\documentclass[twocolumn]{aastex63}

\usepackage{enumitem}
\usepackage{sidecap}
\usepackage{xspace}
\sidecaptionvpos{figure}{t}
\usepackage{appendix}
\usepackage{soul}
\usepackage{rotating}

\newcommand{\x}{X-ray}

\newcommand{\swift}{\textit{Swift}}

\newcommand{\xmm}{\textit{XMM-Newton}}

\newcommand{\nustar}{\textit{NuSTAR}\xspace}

\newcommand{\uhuru}{{\it Uhuru}}
\newcommand{\nicer}{{\it NICER}\xspace}

\newcommand{\ginga}{{\it GINGA}}

\newcommand{\integral}{\textit{INTEGRAL}\xspace}

\newcommand{\exi}{\begin{equation}}
\newcommand{\exo}{\end{equation}}

\newcommand{\straycats}{\texttt{StrayCats}\xspace}
\newcommand{\detone}{\textsc{DET1}\xspace}
\newcommand{\gx}{\mbox{GX~3+1}\xspace}

\accepted{Jan 27, 2021}
\submitjournal{ApJ}

\shorttitle{StrayCats}
\shortauthors{Grefenstette et al.}

\begin{document}

\title{\textit{StrayCats}: A catalog of NuSTAR Stray Light Observations}

\correspondingauthor{Brian Grefenstette}
\email{bwgref@srl.caltech.edu}
\author[0000-0002-1984-2932]{Brian W. Grefenstette}
\affiliation{Cahill Center for Astronomy and Astrophysics, California Institute of Technology, Pasadena, CA 91125, USA}
\author[0000-0002-8961-939X]{Renee~M.~Ludlam}
\altaffiliation{NASA Einstein Fellow}
\affiliation{Cahill Center for Astronomy and Astrophysics, California Institute of Technology, Pasadena, CA 91125, USA}
\author[0000-0002-3669-2294]{Ellen~T.~Thompson}
\affiliation{Space Sciences Laboratory, 7 Gauss Way, University of California, Berkeley, CA 94720-7450, USA}
\author[0000-0003-3828-2448]{Javier~A.~Garc\'ia}
\affil{Cahill Center for Astronomy and Astrophysics, California Institute of Technology, Pasadena, CA 91125, USA}
\affil{Dr. Karl Remeis-Observatory and Erlangen Centre for Astroparticle Physics, Sternwartstr.~7, 96049 Bamberg, Germany}
\author[0000-0002-8548-482X]{Jeremy~Hare}
\altaffiliation{NASA Postdoctoral Program Fellow}
\affiliation{NASA Goddard Space Flight Center, Greenbelt, MD 20771, USA}
\author[0000-0002-3850-6651]{Amruta~D.~Jaodand}
\affiliation{Cahill Center for Astronomy and Astrophysics, California Institute of Technology, Pasadena, CA 91125, USA}
\author[0000-0003-2737-5673]{Roman~A.~Krivonos}
\affiliation{Space Research Institute, Russian Academy of Sciences, Profsoyuznaya 84/32, 117997 Moscow, Russia}
\author[0000-0003-1252-4891]{Kristin~K.~Madsen}
\affiliation{Space Radiation Laboratory, Caltech, 1200 E California Blvd, Pasadena, CA 91125}
\affiliation{Astrophysics Science Division, NASA Goddard Space Flight Center, Greenbelt, MD 20771, USA}
\author[0000-0003-4216-7936]{Guglielmo~Mastroserio}
\affiliation{Cahill Center for Astronomy and Astrophysics, California Institute of Technology, Pasadena, CA 91125, USA}
\author[0000-0002-5752-3780]{Catherine~M.~Slaughter}
\affiliation{Caltech Summer Undergraduate Research Fellowship}
\author[0000-0001-5506-9855]{John~A.~Tomsick}
\affiliation{Space Sciences Laboratory, 7 Gauss Way, University of California, Berkeley, CA 94720-7450, USA}
 \author[0000-0001-9110-2245]{Daniel Wik}
\affil{Department of Physics and Astronomy, University of Utah, 201 James Fletcher Building, Salt Lake City, UT 84112, USA}
\author[0000-0001-9067-3150]{Andreas Zoglauer}
\affiliation{Space Sciences Laboratory, 7 Gauss Way, University of California, Berkeley, CA 94720-7450, USA}



\begin{abstract}


We present \straycats: a catalog of \nustar stray light observations of X-ray sources. Stray light observations arise for sources 1--4$^{\circ}$ away from the telescope pointing direction. At this off-axis angle, X-rays pass through a gap between optics and aperture stop and so do not interact with the X-ray optics but, instead, directly illuminate the \nustar focal plane. We have systematically identified and examined over 1400 potential observations resulting in a catalog of 436 telescope fields and 78 stray light sources that have been identified. The sources identified include historically known persistently bright X-ray sources, X-ray binaries in outburst, pulsars, and Type I X-ray bursters. In this paper we present an overview of the catalog and how we identified the \straycats sources and the analysis techniques required to produce high level science products. Finally, we present a few brief examples of the science quality of these unique data.

\end{abstract}



\section{Introduction} \label{sec:intro}

Compact objects in our galaxy provide an excellent laboratory in which to study matter in extreme conditions. Of most interest are neutron stars (NS) and black holes (BH) in binary systems, where the compact object accretes material from its companion star either through Roche lobe overflow of through a stellar wind from the companion. The inflowing material forms an accretion disk around the compact object with temperatures hot enough to produce copious amounts of thermal X-rays and giving rise to a corona of non-thermal electrons emitting in the hard X-ray band.

\begin{figure*}
  \includegraphics[width=0.5\textwidth]{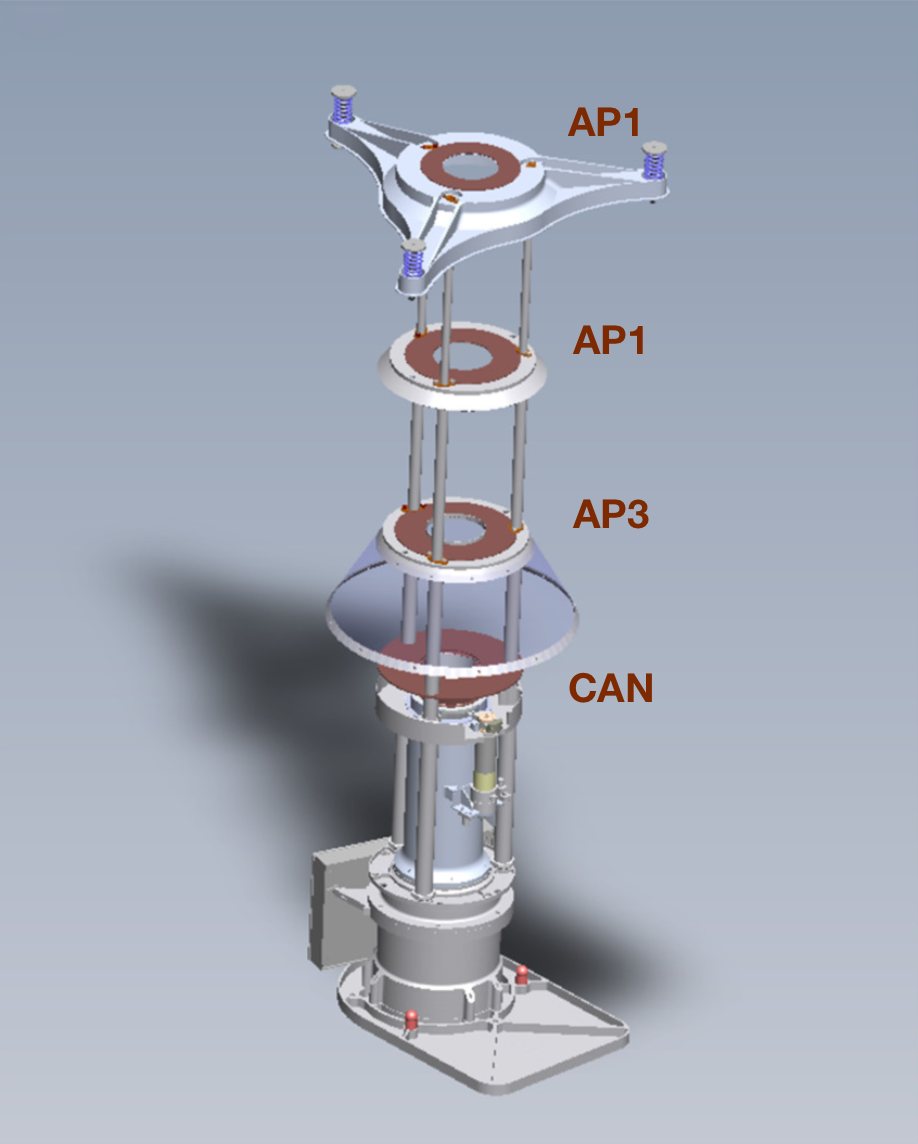}
  \includegraphics[width=0.5\textwidth]{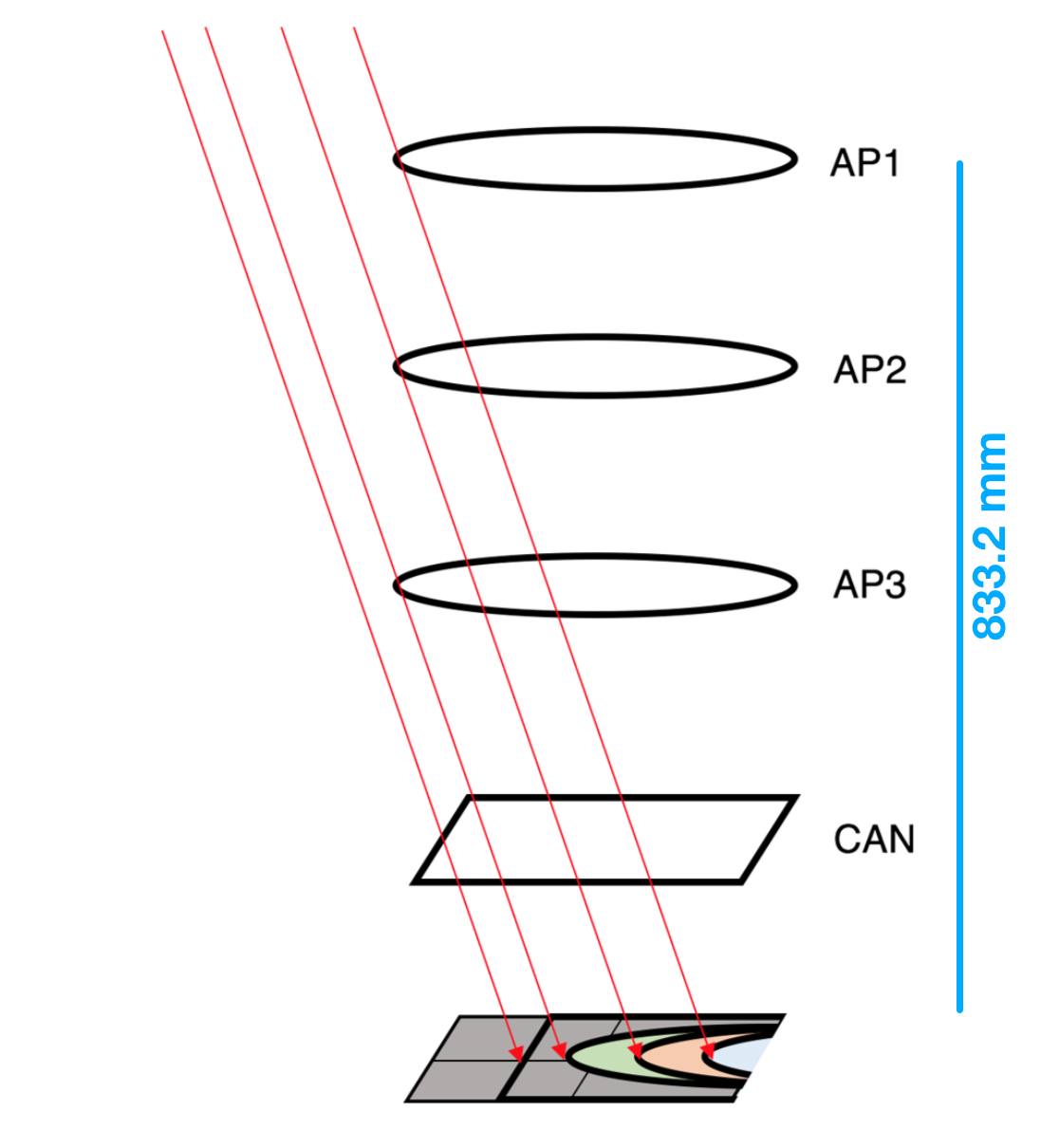} 
   \caption
   { \label{fig:straylight_structures} 
Schematic of the path of stray light photons. (\textit{Left}): CAD rendering of the focal plane and the aperture stop assembly. (\textit{Right}): Red traces show the stray light paths that survive to the focal plane after passing around the aperture stop (AP1, AP2, and AP3) rings and the ``can'' housing the detectors. The height offset from the focal plane to AP1 is shown on the right. Figures adapted from \cite{madsen_observational_2017}.
}
\end{figure*}

The hard X-ray (E$\geq$3 keV) bandpass provides essential diagnostic information on the accretion state of the source and clues to the nature of the compact object in the system. The high energy (E$\geq$20 keV) spectrum of the X-ray binaries in the Galactic plane have been surveyed with low spectral resolution instruments on the \textit{INTErnational Gamma-Ray Astrophysics Laboratory} \citep[\integral,][]{winkler_integral_2003} and the Neil Gehels \textit{Swift} Observatory \citep{gehrels_swift_2004}.

Targeted observations with \nustar \citep[The \textit{Nuclear Spectroscopic Telescope ARray}][]{harrison_nuclear_2013} have demonstrated the diagnostic power of a sensitive instrument over the 3--80 keV bandpass. However, when these sources go into an X-ray bright state they result in extremely high count rates and correspondingly high telemetry loads. Because of this, many observations of bright sources are short in duration ($\approx$ 20-ks) to allow the spacecraft to transmit the data down to the ground without overwriting the storage drives onboard. Unlike \swift, \nustar is not a rapidly slewing instrument, so repeated short monitoring observations of to the same target are not generally possible due to scheduling constraints and require ``Target of Opportunity" programs that can take days or a week to get on target once an observation is trigger.

Fortunately, \nustar can also serendipitously observe bright X-ray binaries through ``stray light." While \nustar is well-known as the first focusing hard X-ray satellite in orbit, the open geometry of the mast that connects the optics to the detectors allows for the possibility of stray light (light that has not been focused by the optics) illuminating detectors. This is typically referred to as ``aperture flux'' since the light passes through the open area of the aperture stops (see Figure \ref{fig:straylight_structures}) and occurs for sources that are roughly 1--4$^{\circ}$ from the center of the \nustar field-of-view \citep{madsen_observational_2017}. 

For most \nustar \ observations, the dominant source of aperture X-ray emission is the cosmic X-ray background (hereafter ``aperture'' CXB, or aCXB). This is the superposition of X-ray light from a uniform background of (unresolved) AGN in the 1--4$^{\circ}$ annulus. This contribution to the \nustar \ background has been well documented \citep[e.g.,][]{wik_nustar_2014} and is generally described by a spatial gradient in the \nustar background across the field of view. 

\begin{figure*}
  \includegraphics[width=0.5\textwidth]{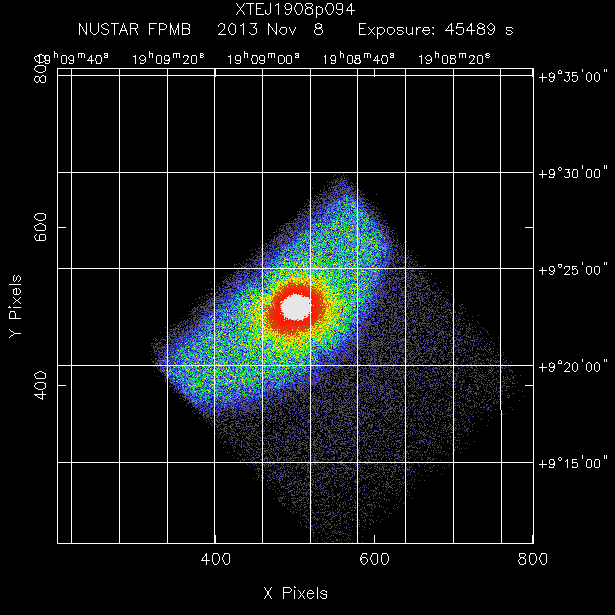}
  \includegraphics[width=0.5\textwidth]{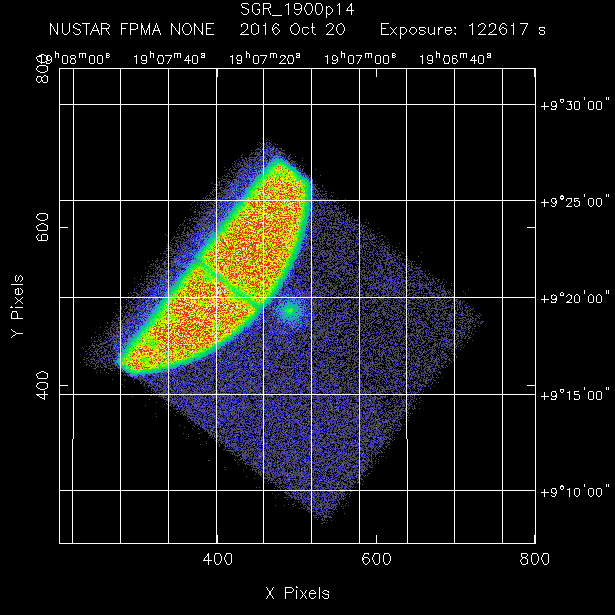}
\caption
   { \label{fig:grs1915} 
3--79 keV \nustar quick look images in ``sky" coordinates from the \textsc{HEASARC} showing the stray light from GRS1915+105 along with the X-rays from the targeted source for two epochs (the \textit{intended} source target name is given in the figure titles). Unlike the point source which is contained on one detector, the stray light spans multiple detectors on the \nustar focal plane.
}
\end{figure*}

When stray light comes from a single off-axis source the emission geometry is much simpler. Instead of a ``gradient'' in the background, we instead observe an easily-identified shadow of the aperture stop ring sharply cutting off the source (Figure \ref{fig:grs1915}). Because the X-rays do not interact with the \nustar \ optics, the response of the instrument is somewhat more straight forward as well. This comes at the reduced effective area for stray light observations compared with pointed observations. 

Recently, observations \textit{intentionally} placing a target so that it is observed via stray light have been undertaken for a number of bright X-ray binaries. This was done to provide contiguous observations while reducing the count rate (and thus the telemetry load) and to potentially extend the spectral range covered by \nustar beyond the 78.4 keV cutoff in the optics response. One example is the observation of the Crab nebula seen via stray light which allows for a simple, unique measurement of the spectral shape and flux of the Crab \citep{madsen_measurement_2017}.

In this paper we describe the \nustar \straycats\footnote{https://nustarstraycats.github.io/}: a catalog of \nustar stray light observations (both serendipitous and intentional) throughout the mission. In \S 2 we describe the preliminary data processing and the stray light identification methodology. In \S 3 we discuss the particular response files needed for \straycats spectroscopic analysis as well as the tools that we have developed for streamlining the extraction of \straycats high level science products, such as spectra and lightcurves. In \S 4 we give an overview of the catalog itself, including source lists and demographics, and in \S 5 we present preliminary analyses of several \straycats data sets to give a demonstration of the type and quality of data. However, we generally will reserve a more detailed follow-up analysis of individual sources to future work.

\section{Data Processing and Stray Light Identification}

Identifying observations contaminated by stray light is non-trivial, due to the variability in the \nustar background contributions, the presence of multiple sources in the field of view (FoV), and the different amounts of detector area illuminated by the stray light sources at different off-axis angles. We utilized two complementary methods: an \textit{a priori} approach based on the location of known bright X-ray sources detected by \swift-BAT and \integral; and a ``bottom up" approach using a statistical approach to identify potential stray light candidate observations.

\subsection{An \textit{a priori} approach}
\label{sec:apriori}

We use the \swift-BAT 105-month all-sky catalog \citep{oh_105-month_2018} of sources along with the \integral 9-year galactic plane ($|b|$~\textless~17.5$^{\circ}$) catalog \citep{krivonos_integral/ibis_2012}. These catalogs are both used by the \nustar Science Operations Center (SOC) to identify and mitigate sources of stray light contamination for science observations. To estimate the amount of stray light in a given observation, we utilize the \texttt{nustar\_stray\_light} IDL code\footnote{https://github.com/NuSTAR/nustar-gen-utils}. This contains a model of the size, shape, and relative positions of the focal plane structures (seen in Fig \ref{fig:straylight_structures}) and the bench that holds the \nustar optics. For a given \nustar pointing orientation and a given stray light target, the ``shadow" from the aperture stop and the optics bench are projected onto the focal plane for each detector to estimate the stray light contribution.

Estimating the strength of the stray light is done by extrapolating the measured spectrum in the \swift-BAT / \integral bands down into the \nustar straylight bandpass (3--20 keV); a process which frequently results in overestimating the \nustar flux for sources that have curvature in the hard X-ray bandpass or have a predominantly thermal spectrum. Nonetheless, there is usually a reasonable match between the brightest catalog sources and the stray light in \nustar.

As a first step, we produce an estimate catalog of all \nustar observations within 4$^{\circ}$ of a ``bright" X-ray source in one of our reference catalogs where we typically define the minimum flux level for a persistent, bright source to be $> 5$ mCrab as measured by the respective instruments on \integral and \swift. This results in several hundred \nustar stray light candidate observations. For each observation we produce the estimated stray light map, and visually compare the results to the observed data. As many of these sources are variable and the internal model of the structures may not be entirely accurate, this does require a human-in-the-loop for positive identification of a stray light candidate. While this process is able to positively identify dozens of stray light observations, it is both inefficient and does not catch any stray light observations of new or intermittently transient sources.

\subsection{A more statistical approach}

Rather than requiring any prior knowledge of a nearby bright target, we instead use the observed data to identify stray light candidates. Since the area of the sky accessible to each \nustar telescope for stray light are different, we treat the two separately.

We first remove contributions from the primary target by first excising all counts from within 3\arcmin\ of the estimated target location. This large exclusion region attempts to account for any astrometric errors between the estimated J2000 coordinates for the target and where the target is actually observed to reduce the ``PSF bleed" from bright primary targets. For bright primary targets (those with focused count rates rates $>$ 100 cps) we find that the primary source dominates over the entire FoV, so we exclude these observations from consideration. Once this is complete, we compute the 3--20 keV count rate for all four detectors on each FPM and combine them to account for the fact that the stray light patterns tend to illuminate one side (or all) of the FoV.

\begin{figure*}
  \includegraphics[width=\textwidth]{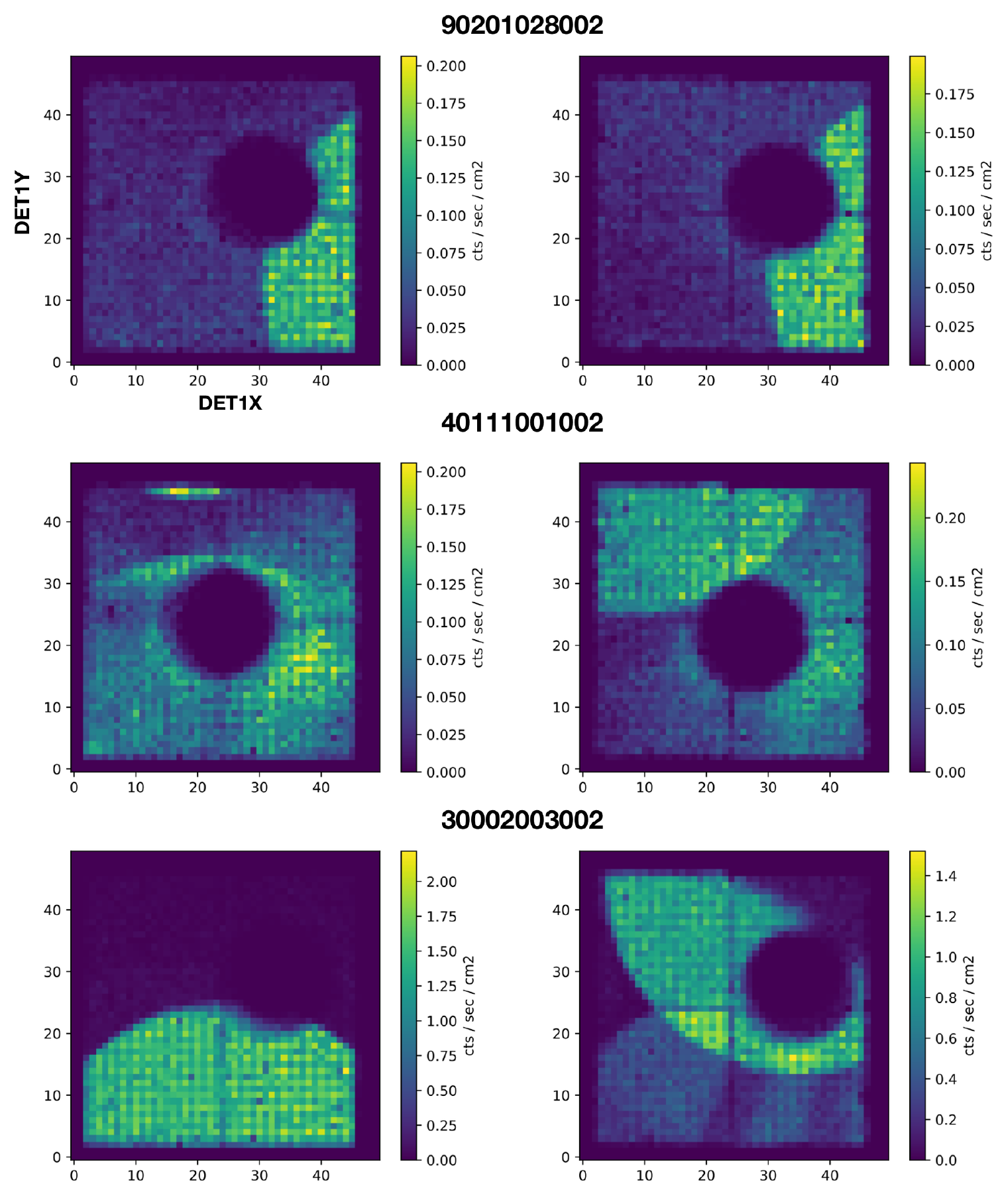}
\caption
   { \label{fig:rogues} 
A rogue's gallery of 3--20 keV \nustar images in \detone pixel coordiantes (1 pixel = 2.54\arcsec = 120.96 $\mu$m) for three \straycats observations showing some of the variety of the stray light patterns in FPMA (\textit{left column}) and FPMB (\textit{right columns}). The primary source has been masked out and the linear colorscale shows the fluence (counts per second per cm$^{2}$) across the field of view for each detector. (\textit{Top}) One of the cases where stray light (here from the LMXB 4U 1624-490) is seen in both FPMs. (\textit{Middle}) A more complex geometry where multiple overlapping or partially blocked stray light sources (the strongest being 4U 1708-2 in FPMA and 4U 1700-377 in FPMB) overlay the extended primary source (RX J1713.7-3946). (\textit{Bottom}) Strong and overlapping stray light from GX 5-1 (lower SL) and GX 3+1 (upper right in FPMB).
}
\end{figure*}

For the remaining sources we flag observations where the count rate measured by a particular detector combination deviates from the mean. Unfortunately, due to extended sources, fields with multiple point sources, and intrinsic variation in the \nustar background, all of the candidate \straycats observations had to be further checked by eye. We do this by constructing \detone images in the 3--20 keV bandpass and look for the signatures of stray light. Figure \ref{fig:rogues} shows a selection of \straycats observations where the SL can clearly be seen. 

We continue the iterative process to identify candidates described above until all of the candidates appear to be simply variations in the \nustar background and not clearly associated with stray light. Overall, more than 1400 candidate stray light observations were checked by hand for the presence of stray light.

We feel confident that we have thus identified all of the stray light sources that could (a) produce a strong enough signal to impact science analysis of the primary target and (b) be useful for scientific analysis in their own right. These fully vetted \straycats sources form the basis for the full catalog. In addition to stray light, we have also identified a number of observations where targets just outside of the \nustar FoV result in ``ghost rays", where photons perform a single-bounce photons off of the \nustar optics rather than the double-bounce for focused emission \citep{madsen_observational_2017}. These are included in \straycats for completeness. 

We do note that this human-in-the-loop approach does result in a bias where faint stray light sources are more easily seen during long exposures. Similarly, sources with transient flaring behavior on timescales of a few 100-$s$ will be difficult to identify unless the quiescent flux level is greater than that of the standard \nustar background. We anticipate that a further investigation for transients could produce a number of additional \straycats candidates, though this is beyond the scope of this first work.

\section{The \straycats catalog}
\label{sec:catalog}

The \straycats Catalog is intended to be used by observers looking for serendipitous observations of bright galactic (including the LMC and SMC) sources beyond what is available through traditional monitoring observations. The catalog is available via a simple web interface \footnote{https://nustarstraycats.github.io/} or simply through a FITS file that identifies which \nustar sequence IDs contain \straycats sources. For observations that contain multiple \straycats sources the web interface also contains diagnostic information that can be used to determine which stray light pattern is associated with a particular source (i.e., the images shown in Fig \ref{fig:rogues}). An excerpt of the table is given in the Appendix in Table \ref{tab:straycats}. \\

The first version of \straycats includes the following columns:
\begin{itemize}

    \item StrayID: The \straycats catalog identifier, which is StrayCatsI\_XX where XX is the row number after the catalog is sorted the RA and Dec for the \nustar sequence ID.
    
    \item Classification
    \begin{enumerate}
        \item SL: The source has been positively identified as a \straycats target
        \item Complex: Stray light is present, but there are multiple overlapping stray light regions that make the sources difficult to identify
        \item Faint: Stray light is present, but is too faint to be positively identified.
        \item GR: The observation contains ghost-rays from sources just outside of the FoV
        \item Unkn: A stray light pattern is present, but the source of the stray light remains unknown.
    \end{enumerate}

    \item SEQID: The \nustar sequence ID

    \item Module: The \nustar FPM that contains the stray light (A or B)

    \item Exposure: The exposure time for this observation in seconds

    \item Multi: Whether the sequence ID contains multiple stray light patterns (Y or N)
    
    \item Primary: The name of the primary target for the pointed science observation

    \item TIME / END\_TIME: The MJD start/end of the observation
    
    \item RA/DEC\_Primary: RA/Dec of the \textit{primary} target

    \item SL Source: The name of the source of SL if we have identified it
    
    \item SL Type
    
    For sources with a positive identification, we have made an effort to sample the literature and provide a source classification. Many of these are relatively famous sources identified by \ginga\ or \uhuru\ with a large literature background, so we do not provide prime references for the classifications in \straycats. For sources with Classification other than SL, this defaults to ``??". Classification types are:
    
    \begin{enumerate}
        \item AGN: Active Galaxy
        \item LMXB (low-mass X-ray binary) with -NS or -BH if the compact object type is known
        \item HMXB (high-mass X-ray binary) with -NS or -BH if the compact object type is known
        \item Pulsar / PWNe (Pulsar Wind Nebula) / NS
        \item BHC (Black Hole Candidate)
        \item SNR (Supernova Remnant)
        \item Cluster (Galaxy cluster)
        \item Radio Galaxy
    \end{enumerate}
    
    \item SIMBAD\_ID: The identifier that can be used via SIMBAD to identify the source. This can often be different
    than the source name in the all-sky catalogs used to identify the source (if known, otherwise defaults to NA)
    
    \item RA/DEC\_SL: RA/Dec of the source of the stray light (if known, otherwise defaults to -999).

\end{itemize}

\straycats contains 436 telescope fields (with A and B counted separately) containing stray light from 78 confirmed \straycats sources. During the visual inspection of the stray light candidates, we compare the observed stray light patterns with those predicted for that observation using the same code used in \S \ref{sec:apriori}. For a majority of sources, this is sufficient to identify the source of stray light. For a few dozen cases, the stray light is associated with a source \textit{not} present in either catalog. This was either because the source was a new transient (e.g., a number of MAXI-identified transients that went into outburst over the last few years), the source is only occasionally detected by the all-sky hard X-ray detectors (e.g., sources contained in the ``\swift-BAT historically detected" list), or the source is typically too soft to be detected by \swift-BAT or \integral. We have not yet identified any previously unknown \straycats sources.

\begin{figure}[ht]
\begin{center}
  \includegraphics[width=0.5\textwidth]{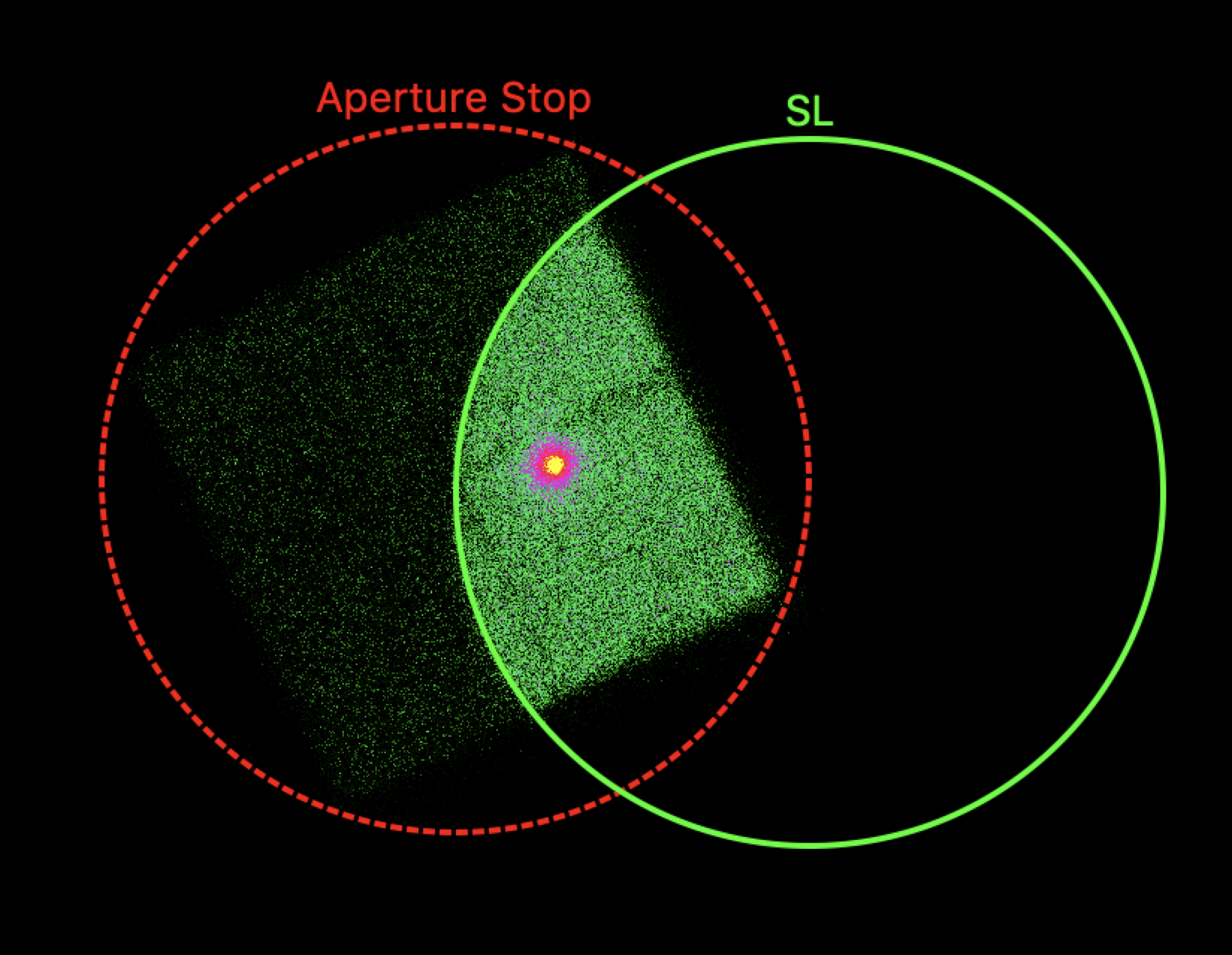}
   \caption{ \label{fig:cirx1} 
An example of the ``stray light" (SL, green) region and the ``aperture stop" region (red, dashed) that can be used to identify the source location on the sky. See text for details. 
}
\end{center}
\end{figure}

We can esimate the source location using the projected shape of the aperture stop on the focal plane. Fig \ref{fig:cirx1} gives an example of this for a simple case. Here, the curvature of the aperture stop shadow is clearly seen on the focal plane. We generate a ``SL" region that matches the known curve, and compute the offset between this and the center of the FoV (the ``Aperture Stop" region in Fig \ref{fig:cirx1}). We can compute the offset on the focal plane (in mm) and leverage the fact that we know that the deployed aperture stop is 833.2 mm (Fiona Harrison, priv comm.) away from the focal plane to convert this offset to angular offset. The direction of the shift (in sky coordinates) allows us to determine the position angle of the shift. In the example shown here, we were able to reproduce the location of Cir X-1 to better than 10\arcmin, which is generally good enough to identify the source. For cases where multiple overlapping stray light patterns are seen and we cannot unambiguously identify the source we assign the ``Complex" classification pending a detailed analysis.

\begin{figure*}[ht]
  \includegraphics[width=\textwidth]{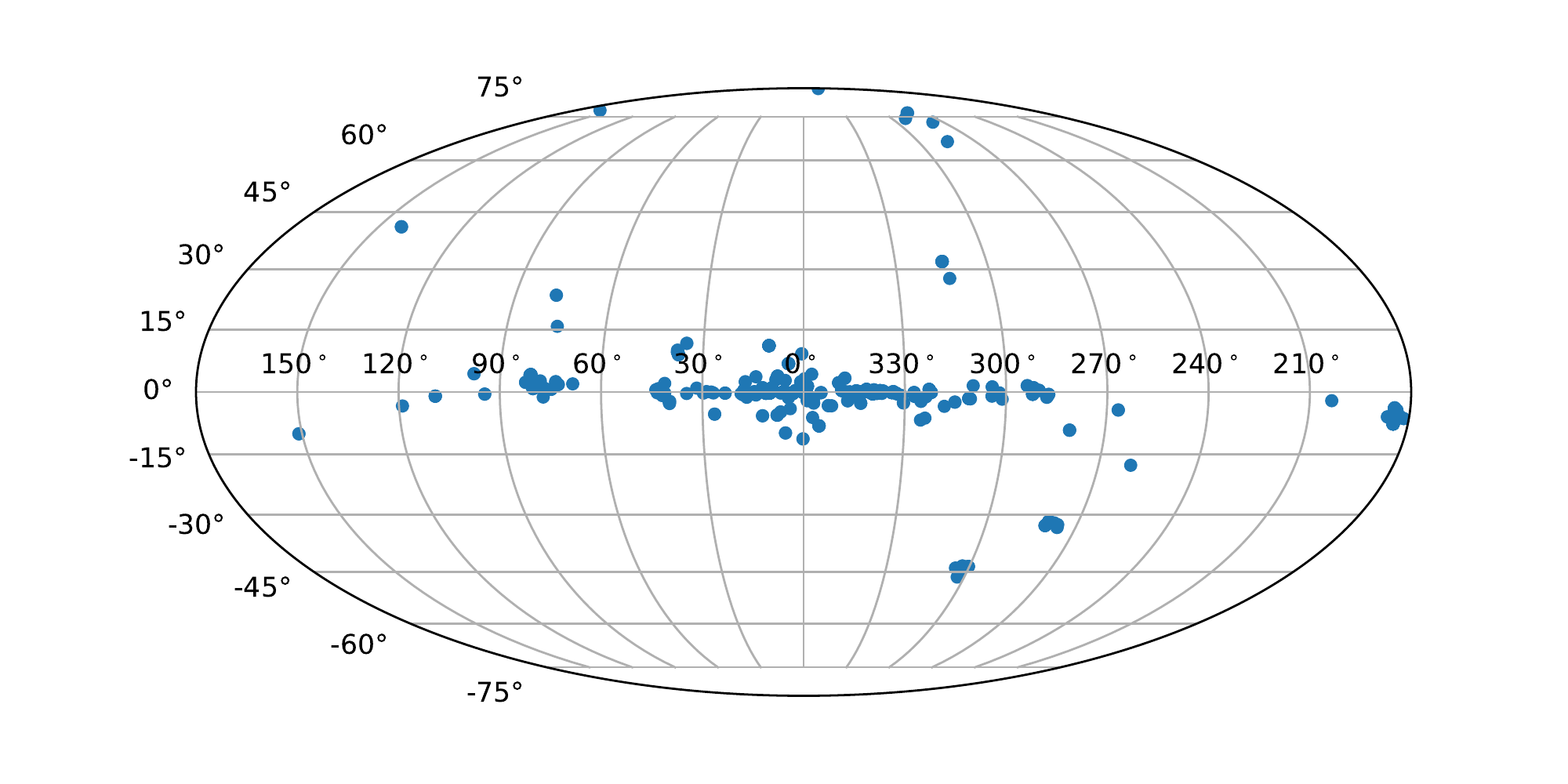}
   \caption{ \label{fig:straylight_galatic} 
Distribution of the \straycats in galactic coordinates showing the clustering of these sources near the Galactic plane, the contribution from bright sources in the LMC and SMC, and a few AGN located out of the plane of the Galaxy. The coordinates shown here are the for the primary (focused target).
}
\end{figure*}

The catalog contains seven AGN and one galaxy cluster, several pulsar wind nebulae and supernova remnants, roughly 17 accreting black holes (including black hole candidates), as well as over forty accreting neutron stars including several pulsars and a number of known Type I X-ray bursters. Figure \ref{fig:straylight_galatic} shows the galactic distribution of these sources, where the density of sources near the galactic plane and the LMC and SMC can clearly be seen.

\section{\straycats data analysis tools and response files}

\straycats require subtly different analysis methods than those typically used for focused \nustar observations. Rather than working in ``SKY" coordinates like focused observations, for stray light observations we instead work in ``detector'' coordinates (\detone coordinates in \nustar vernacular). This coordinate system is fixed with respect to the \nustar CdZnTe detectors and, in these coordinates, the pattern of stray light on the focal plane is predominantly sensitive to the observatory orientation and is extremely weakly coupled to any motion of the \nustar mast. For pointed observations, the $\sim$mm-scale motion of the \nustar mast affects the throughput of the optics by changing the distance of the source from the optical axis \citep[``vignetting"][]{harrison_nuclear_2013}. In non-focused observations the mast motion only minimally changes the shadow pattern as observed by the detectors and can be neglected. 

Producing high-level science products for a \straycats observation is relatively straightforward. These mostly deal with properly tracking the production of ``source" regions files and applying spatial filtering on the \nustar data in \detone coordinates. Our goal is to make the resulting products as similar to standard \nustar products as possible for the ease of use.

To date, we have contributed a number of high-level ``wrappers" to the \nustar community-contributed GitHub page\footnote{https://github.com/NuSTAR/nustar-gen-utils}. These are largely written in python and significantly leverage the existing astropy framework \citep{robitaille_astropy_2013, astropy_collaboration_astropy_2018}, as well as the multi-mission FTOOLs distributed by the HEASARC, such as \textsc{XSELECT}. Final high-level products are mostly generated using \textsc{nuproducts} from the \textsc{NuSTARDAS} software with a number of non-standard configuration settings. This allows a user to easily produce standard spectrum (PHA) and lightcurve files as well as response matrix functions (RMFs) which can directly be loaded into downstream analysis software such as \textsc{Xspec} \citep{arnaud_xspec:_1996} or \textsc{ISIS} \citep{houck_isis_2000} for spectral analysis or \textsc{Stingray} \citep{huppenkothen_stingray_2019} for timing analysis.
\subsection{Response Files}
The one unique requirement for the analysis of \straycats observations is the production of the response files. For a focused observation, each count is first ``projected'' onto the sky and the optics response (i.e. the ancillary response file, or ARF) is produced so that it accounts for the time-dependent drift in the location of the optical axis due to the thermal motion of the \nustar mast. The ARF is generated starting with an on-axis optics response, which is then convolved with energy-dependent vignetting function based on the off-axis angles sampled by the source. Finally, the ARF also includes the attenuation along the photon path due to the optics thermal covers, the Be window protecting the detectors, and the absorption features in the CdZnTe detectors themselves\footnote{see the NuSTAR software user's guide: https://heasarc.gsfc.nasa.gov/docs/nustar/analysis/nustar\_swguide.pdf}.

Since for \straycats  observations we are working in \detone coordinates, we no longer need to account for the time-dependent variations in the ARF, nor (obviously) the response of the optics themselves. The \straycats  ARF, instead, only needs to account for the amount of illuminated area on the focal plane (for overall normalization, given in cm$^2$) and any energy-dependent absorption due to the Be window and losses in the CdZnTe detectors. All of these contributions are currently stored in the \nustar CALDB files (with the exception of the Be window attenuation, which is subsumed into the on-axis ARF in the CALDB). The ARF generation tool for \straycats  analysis properly reads these files from the \nustar CALDB and weights the response based on the illuminated area on each focal plane detector. The resulting file can be directly imported into \textsc{XSPEC} along with the other spectral files above for analysis. This approach has been validatd against observations of the Crab \citep{madsen_measurement_2017}.

Absorbed stray light \citep[stray light that partially penetrates through the aperture stops, see][]{madsen_observational_2017} is not accounted for here. These response files only account for the unabsorbed stray light that reaches the focal plane. In addition, two of the sources in \straycats are extended sources (Cas A and the Coma Cluster). Analyzing data from extended sources is more complex and beyond the scope of this analysis. Analyzing these sources in detail will likely require bespoke ray-trace simulations to properly interpret the stray light spectrum.

\subsection{Region Files}

While all of the \straycats clearly show the effects of stray light, the scientific usefulness of the observations will depend on how much of the FoV is covered by stray light. In the case of the intentional stray light observations mentioned above, the \nustar observations was designed to maximize the amount of detector area illuminated by stray light, which results in roughly half of the 16 cm$^2$ detector area being illuminated (compared with the on-axis effective area of $\approx$400 cm$^2$ for each \nustar telescope). For standard observations, the \nustar SOC attempts to minimize this coverage when possible, so the illuminated detector area for the serendipitous \straycats observations varies dramatically. Because the stray light pattern depends on the shadowing of the detectors by the optical bench, \nustar is also rarely in an orientation where stray light is present on \textit{both} \nustar FPMs. 

Due to the large number of \straycats, and the geometrically complex region shapes, we developed a semi-automated approach to reduce the amount of manual effort involved in generating the optimal extraction region. The “wrapper” for this approach is available in the aforementioned \nustar GitHub page. For \straycats containing a bright point source the first step of this process is point source removal. This is done first by determining the position of the \textit{targeted} source in \detone coordinates (using the \texttt{nuskytodet} FTOOL). This location depends on the motion of the \nustar mast and any changes in the \nustar pointing, so we determine the radial distance from each observation count from this position. We screen events within $r$-arcminutes of the source (if necessary, and where the choice of $r$ is chosen on a case-by-case basis) and generate an image in an adjustable energy band (the 3--10 keV band is default). 
\begin{figure}[]
\begin{center}
  \includegraphics[width=0.5\textwidth]{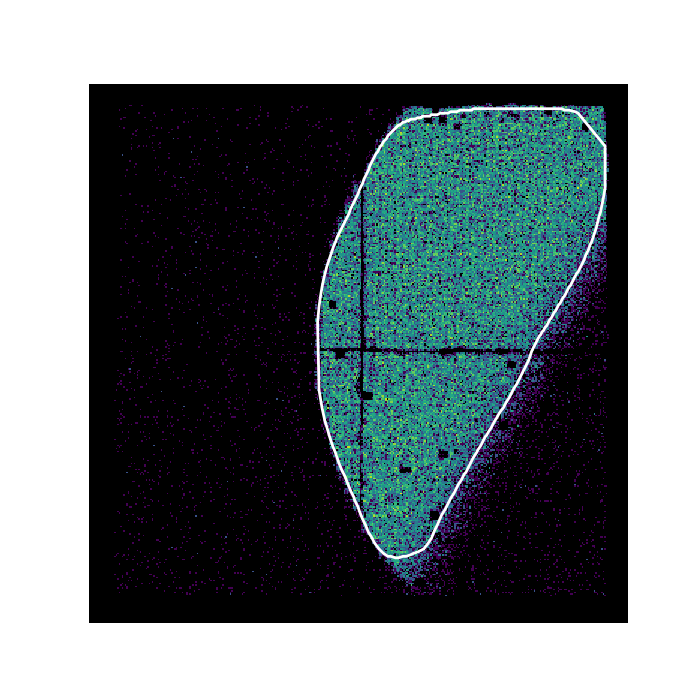}
   \caption{ \label{fig:autoregion} 
Example of a semi-automatically-generated region for a \straycats  observation of the Crab.
}
\end{center}
\end{figure}

We use Canny edge detection from \texttt{scikit-image}\footnote{https://scikit-image.org} to generate the polygons used to estimate the source region where the width of the Gaussian filter used by the Canny edge detection ($\sigma$) is an adjustable parameter. Again, this is chosen on a case-by-case basis such that the filter accurately identifies the edges of the stray light region. Polygon region corners in image coordinates are determined from the detected edge pixels and used to write a region file in SAOImageDS9 standard format using the \texttt{regions} astropy-affiliated module. 

This approach is particularly useful for stray light regions with an angular cutaway resulting from the shadow of the optical bench (i.e., Fig \ref{fig:autoregion}). This process is most efficient for intentional stray light observations and serendipitous observations containing a single stray light pattern only from the ``SL Target" source (i.e., entries in the \straycats catalog with the Classification ``SL” and Multi value ``N”). Currently, this approach is most limited by the $\sigma$ parameter, which approximately ranges between 3 and 12 for optimal stray light observations but can vary greatly for weak stray light regions. Discontinuities in the edges identified by the Canny filter occasionally result in the created polygon region omitting (sometimes negligibly thin) slices of the stray light region; these anomalies can often be corrected by fine-tuning $\sigma$. However, there are no optimal $\sigma$ values for the Canny filter to properly identify the stray light region for observations in which the fluxes of the background and the stray light are comparable. Future improvements to this process that eliminate the manual determination of the point source removal limit and Canny edge detector sigma would allow for fully-automated region extraction.

\subsection{Background}

Dealing with background for \straycats  sources is not trivial. For standard \nustar pointed obervations, standard techniques such as using a neighboring source-free region to estimate the background and/or estimating the \nustar background through tools such as \texttt{nuskybgd} \citep{wik_nustar_2014} can be used ``out of the box". However, as we are using \nustar as a collimator rather than a focusing telescope, the background must be treated with more care.

The \straycats  source regions cover a large region of the FoV (and there may be multiple \straycats  sources as well as the primary source in the FoV), so selecting a background region may be difficult. In addition, for bright \straycats  sources, some stray light may also be transmitted through the aperture stop at higher energies, making it impractical to select a neighboring ``source free" region of the FoV to use to estimate the backgrounud \citep[see][for further discussion]{madsen_observational_2017, madsen_measurement_2017}.

\begin{figure*}[ht!]
  \includegraphics[width=1.0\textwidth]{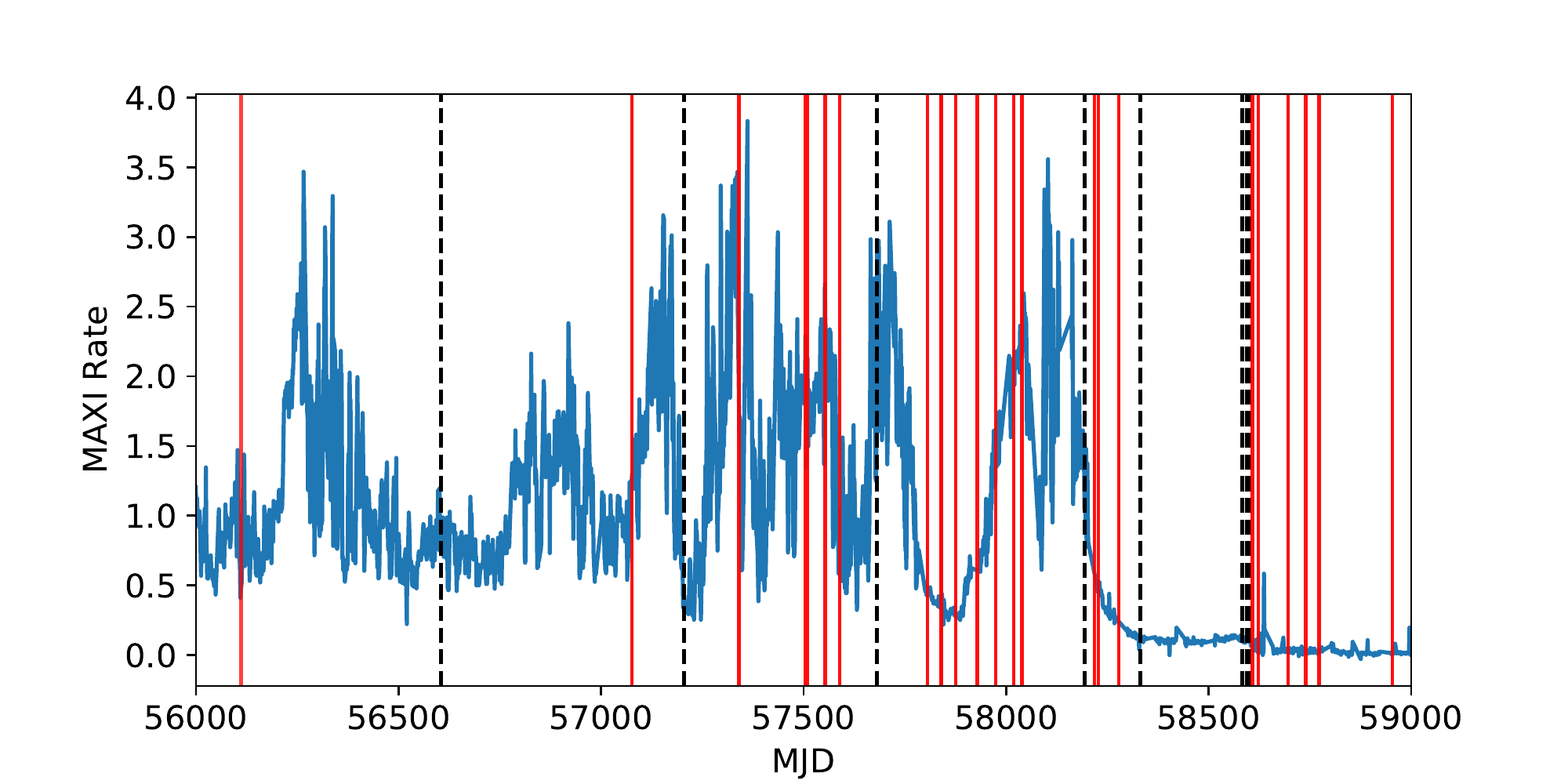}
   \caption{ \label{fig:grs1915longterm} 
The long-term 2-10 keV lightcurve of GRS 1915+105 as measured by MAXI (blue histogram) along with the timing of focused \nustar observations (red lines) and the \straycats  observations (dashed black lines). The final three epochs are clustered in the 16 days just before MJD 58600.
}

\end{figure*}

Modeling the background contributions also must be handled with care. Because many of the \straycats  sources are near the Galactic plane, the standard models of the spatial variation of the \nustar background used by \texttt{nuskybgd} to model the contributions from the Galactic ridge X-ray emission \citep[][GRXE]{krivonos_hard_2007} are largely untested and may need to be adapted for the non-isotropic shape of the GRXE.

The exact method used to handle the presence of background will necessarily vary depending on the science goals for the individual analysis. For bright, hard sources, even without the aid of the \nustar optics, the backgrounds in \nustar are so low that the background may be neglected up to high energies. For fainter sources (or soft sources) the energy at which the background starts to significantly contribute (and therefore the background component which matters the most for spectral analysis) will depend on the details of the source flux. We do not expect there to be a universal solution or recommendation for how to handle the backgrounds.

In the selected preliminary results below, spectral analysis is typically halted when the source flux falls so that the background is estimated to be $\sim$10\% of the source flux, but we stress that a thorough treatment of the background must be considered.

\section{Selected preliminary \straycats  results}

\begin{figure}[ht!]
\begin{center}
  \includegraphics[width=0.5\textwidth]{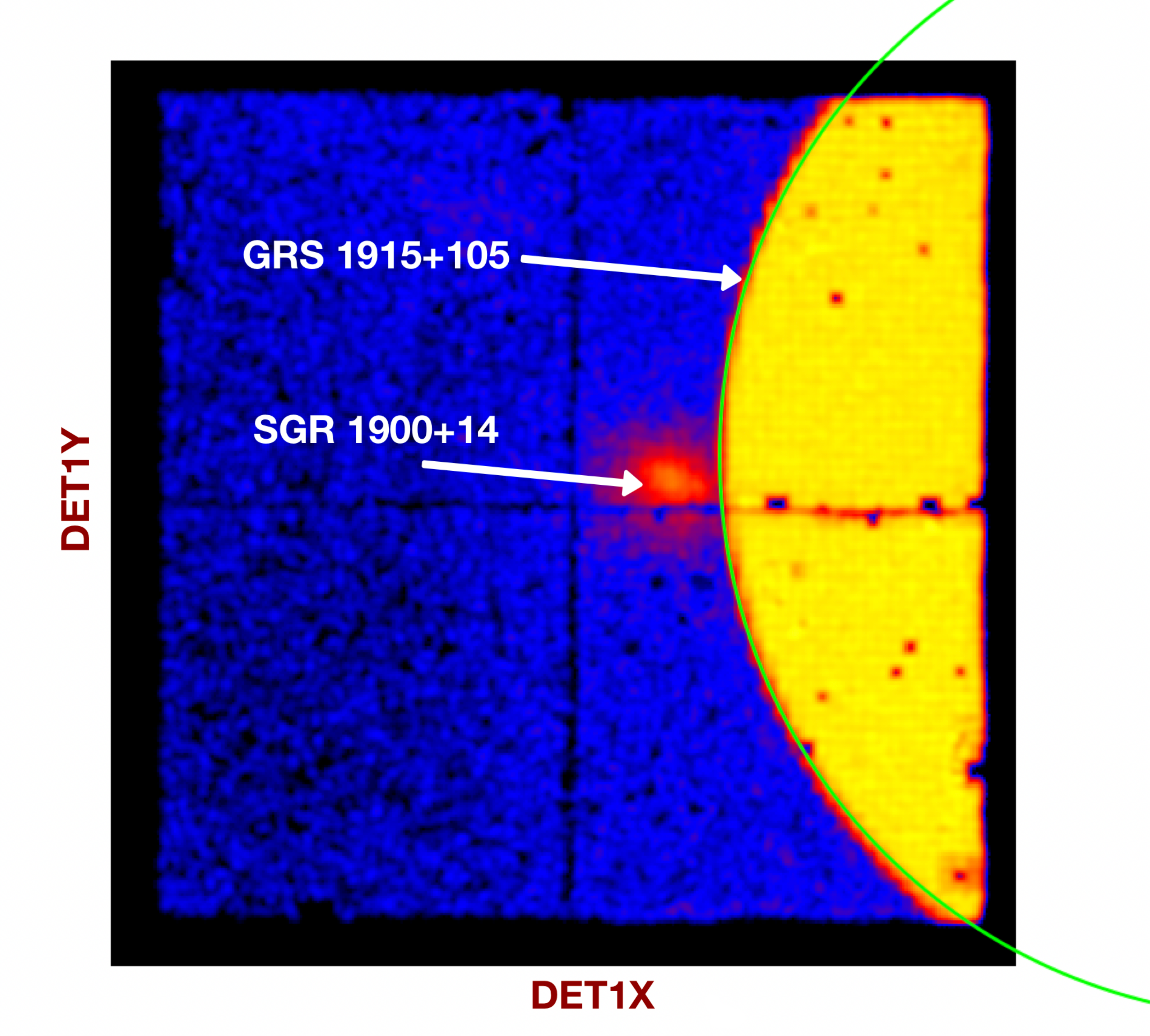}
     \caption{ \label{fig:grs1915im}  The 3-20 keV \detone image for sequence ID 30201013002. The faint primary source is shown, as is the stray light pattern for GRS 1915+105 along with the region showing the shadow of the aperture stop.}
\end{center}
\end{figure}

\begin{figure}[ht!]
\begin{center}
  \includegraphics[width=0.45\textwidth]{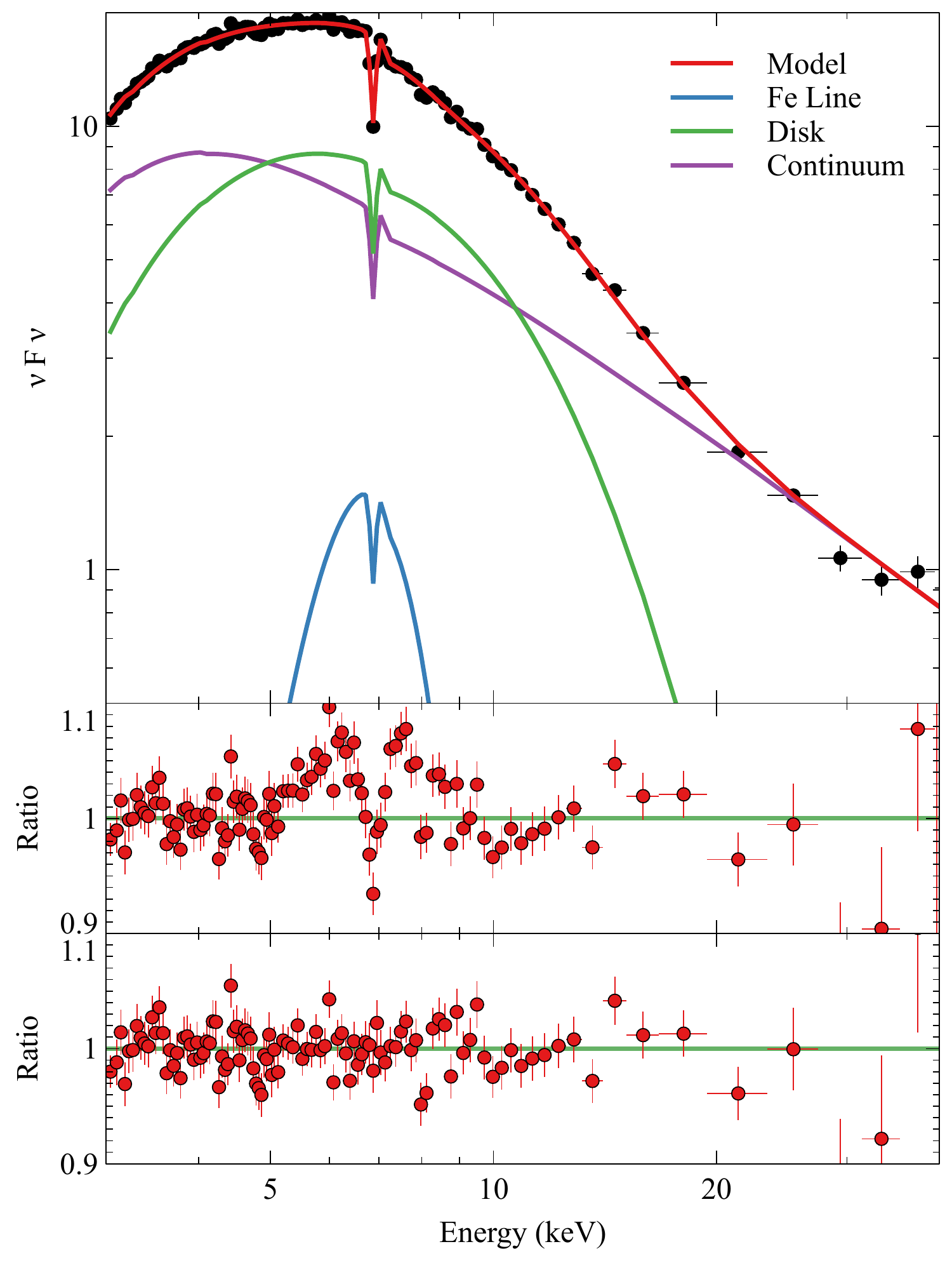}
     \caption{ \label{fig:grs1915spec}  The integrated spectrum from Obs3A from a portion of the stray light region. A similar sized region was used to estimate the background. The base model spectrum here consists of a hot accretion disk component and a soft non-thermal power-law, though this leaves strong residuals near the Fe line (\textit{middle}). We find that after the addition of a broad Fe line and absorption features associated with disk winds in this system that we obtain a reasonable fit to the data.
     }
\end{center}
\end{figure}

\subsection{GRS 1915+105}

\begin{table*}[htb!]
\caption{GRS 1915+105 \straycats  Observations}
\label{tab:grsobs}
\begin{center}
\begin{tabular}{lccccccc}
\hline
Obs \# & Sequence ID & Obs.\ Date &  (MJD) & FPM & Exp.\ (ks) & area (cm$^{2}$) \\
\hline
1A & 80001014002 & 2013-11-08T18:11:07 & 56604.8 & A &   45.15 & 3.9 & \\
1B  & - & - & - & B &   45.49 & 4.0 &\\
2 & 30101050002 & 2015-07-01T15:31:08 & 57204.6 & A &    41.34 & ** &\\
3A* & 30201013002 & 2016-10-20T16:56:08 & 57681.7 & A &   122.3 & 2.2 & \\
3B & - & - & - & B &   122.6 & 3.6 & \\
4 & 40301001002 & 2018-03-17T01:46:09 & 58194.1 & A &   125.6 & 5.9 & \\
5 & 30401018002 & 2018-08-01T12:41:09 & 58331.5 & B &    78.3 & 4.9 & \\
6 & 90501317002 & 2019-04-10T01:26:09 & 58583.1 & A &    40.8 & 5.7 & \\
7 & 30402026002 & 2019-04-22T00:11:09 & 58595.0 & A &    18.83 & ** & \\
8 & 30402026004 & 2019-04-26T13:41:09 & 58599.6 & A &    23.31 & ** & \\
\hline
\end{tabular}
\end{center}
*:Used for the analysis in this work; **:Small stray light area
\end{table*}

\begin{figure}[htb!]
\begin{center}
  \includegraphics[width=0.5\textwidth]{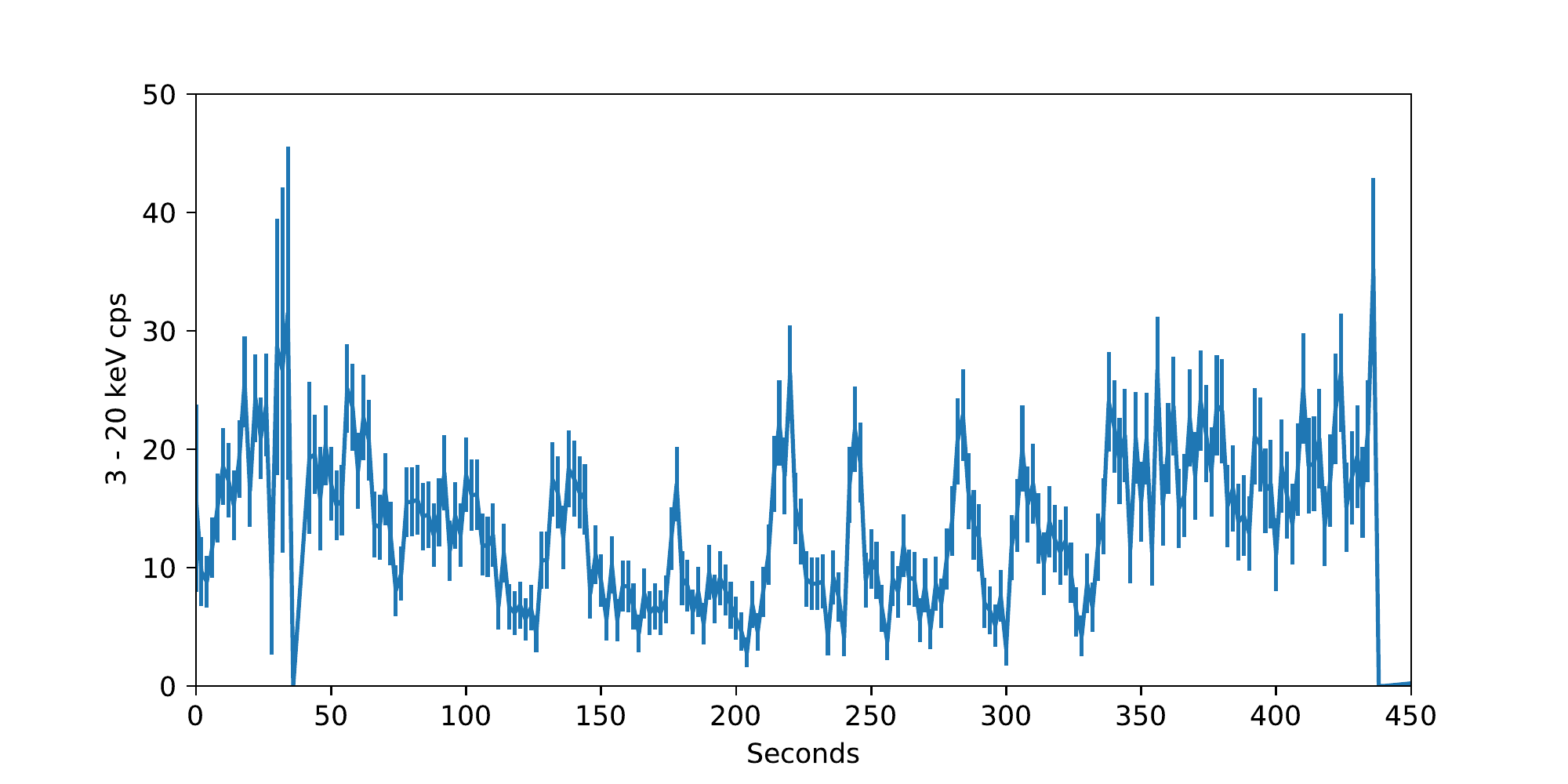} \\ \includegraphics[width=0.5\textwidth]{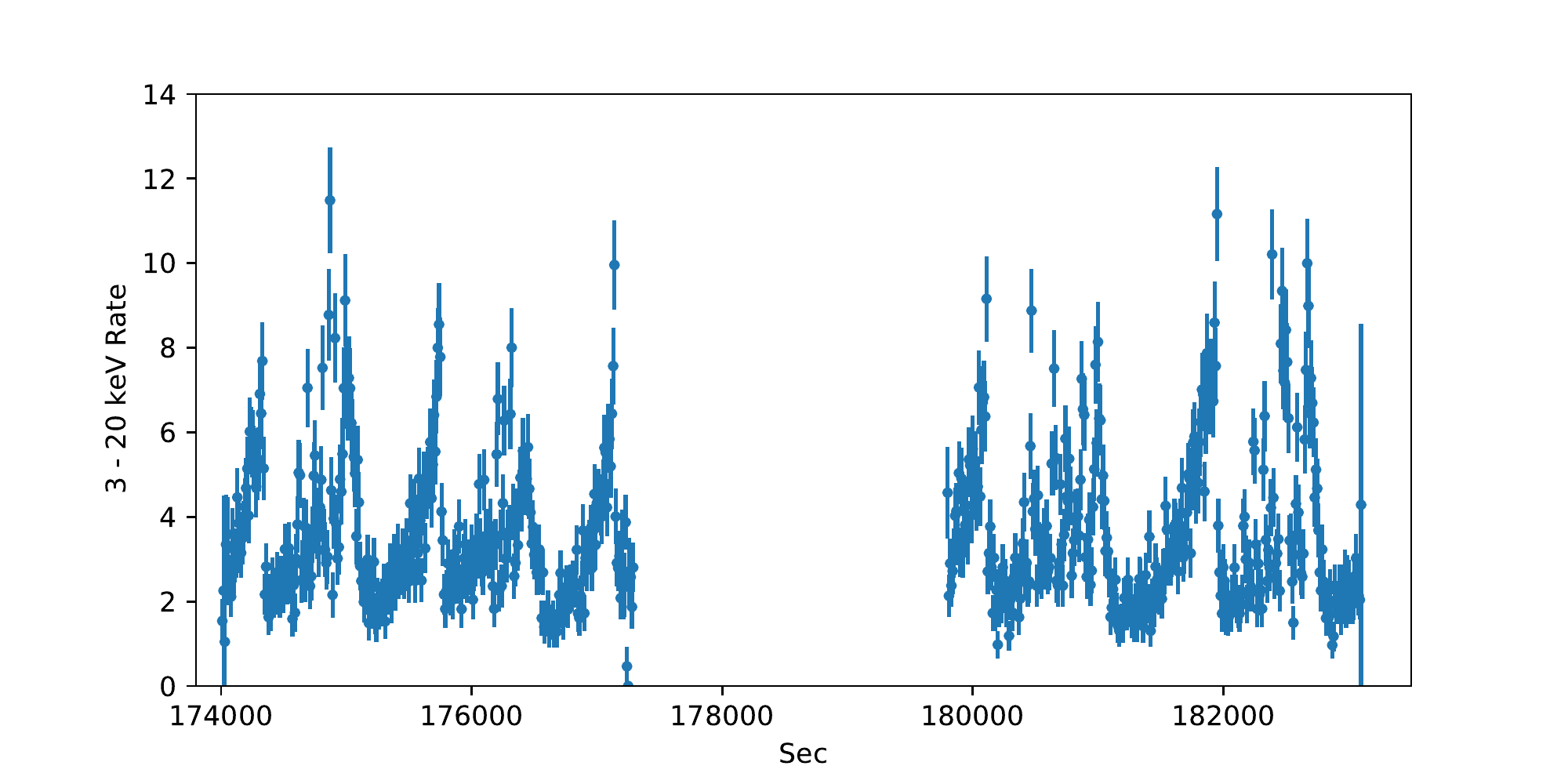}     \caption{ \label{fig:grs1915lc}  The 3-20 keV lightcurve for the first 450-s of the first orbit, binned at 2-s resolution shows the presence of transient slow (mHz) QPO signals. (\textit{Bottom}) The 3-20 keV of two later orbits binned at 10-s resolution showing that the source has transitioned to its $\theta$-state.}
\end{center}
\end{figure}

GRS 1915+105 is a LMXB system which has been in outburst since its discovery in 1992 \citep{castro-tirado_grs_1992} and shows a wide range of source spectral and timing states \citep[e.g.,][]{belloni_model-independent_2000}. The system is known to host a near-maximally spinning black hole \citep{mcclintock_spin_2006} and observations of the absorption features also reveal the presence of a complex outflowing disk wind \citep{miller_accretion_2016}. However, the source began a decay to either a quiescent or a highly absorbed state between 2018 and 2020 \citep{miller_obscured_2020, neilsen_nicer_2020}. Since 2012, \nustar has observed the source a number of times at varying flux levels (Fig \ref{fig:grs1915longterm}). However, the high count rates from this source present two key problems that affect the scientific return from these data: (1) \nustar has a fixed 2.5-ms deadtime-per-event, resulting in a maximum throughput of 400 cts~s$^{-1}$. In high rate sources this deadtime also results in the effective exposure being much lower than the time spent observing the target; (2) As mentioned above, the high count rates result in high telemetry loads that require short duration observations to avoid data loss on board. GRS 1915+105 also appears in 6 \straycats epochs, covering a wide range of flux states (Figure \ref{fig:grs1915lc}) as measured by the Monitor of All-sky X-ray Image (MAXI) instrument on the \textit{International Space Station} \citep{matsuoka_maxi_2009}. The duration of the \straycats  observations vary, with several snapshots roughly 20-ks effective exposure to several deep observations with over 120-ks of exposure. A summary of the \straycats for GRS 1915+105 is given in Table \ref{tab:grsobs}.

\begin{table*}[htb!]
\caption{GX~3+1 \straycats  Observations}
\label{tab:obs}
\begin{center}
\begin{tabular}{lccccc}
\hline
Obs \# & Sequence ID & Obs.\ Date & FPM & Exp.\ (ks) & area (cm$^{2}$) \\
\hline
1 & 30002003003 & 2013-06-19T09:31:07 & B & $\sim29$ & 3.51  \\
2 & 80002017002 & 2014-02-15T05:36:07 & A & $\sim39$ & 4.64  \\
3 & 90101012002 & 2015-08-11T22:51:08 & B & $\sim49$ & 1.46  \\
4 & 90101022002 & 2016-02-18T22:26:08 & A & $\sim36.7$ & 3.79  \\
5 & 40112003002 & 2016-03-17T00:31:08 & A & $\sim52$ & 1.35  \\
6 & 80102101002 & 2016-09-29T21:21:08 & B & $\sim29.5$ & 6.33 \\
7 & 80102101004 & 2016-10-19T15:01:08 & B & $\sim28$ & 7.15  \\
8 & 80102101005 & 2016-10-31T20:11:08 & B & $\sim29$ & 6.66  \\
9 & 80202027002 & 2017-02-18T14:31:09 & A & $\sim31$ & 4.69  \\
10* & 40112002002 & 2017-04-03T18:31:09 & A & $\sim100.7$ & 4.18 \\
11 & 90402313004 & 2018-04-14T02:56:09 & A & $\sim61$ & 3.40\\
 &  - & - & B & $\sim61$ & 3.43 \\
12 & 90501329001 & 2019-06-22T07:51:09 & B & $\sim 40$ & 3.35 \\
13 & 90501343002 & 2019-10-01T22:36:09 & B & $\sim37$ & 1.65  \\
14 & 90601317002 & 2020-05-07T07:06:09 & A & $\sim49$ & 4.12\\
\hline
\multicolumn{3}{l}{*:Used for the analysis in this work}
\end{tabular}
\end{center}
\end{table*}

As an example, we show preliminary results from one epoch (Obs 3A, 30201013002, Figure \ref{fig:grs1915im}), which had an effective exposure of 122 ks spanning over roughly 240 ks (over two and a half days) of clock time. The epoch-averaged source spectrum (Figure \ref{fig:grs1915spec}) shows that the source is clearly detected up to at least 40 keV before background becomes a significant contribution to the spectrum. At low energies we clearly see evidence for a Fe-line features and absorption features typically associated with disk winds in this system \citep[e.g,][]{miller_accretion_2016, neilsen_persistent_2018}.

However, the spectrum for this source is known to be highly variable with the source hardness varying with the apparent emission states and throughout this extended observation the source showed a variety of emission states. For example, during the first orbit we clearly observe QPOs in the form of 10 to 20-$s$ recurrent ``pulsations" of emission, while in later orbits during the same observation the source has transitioned to its $\theta$-state, showing emission building up over the span of a few hundred seconds before sharply dropping away (Fig \ref{fig:grs1915lc}). A detailed analysis of the spectral changes throughout this system is beyond the scope of this work \citep[e.g.,][]{zoghbi_disk-wind_2016}, but shows the utility of only one of the several observations of GRS 1915+105.

\begin{figure}[!]
\begin{center}
\includegraphics[width=0.5\textwidth,trim=5 0 0 0,clip]{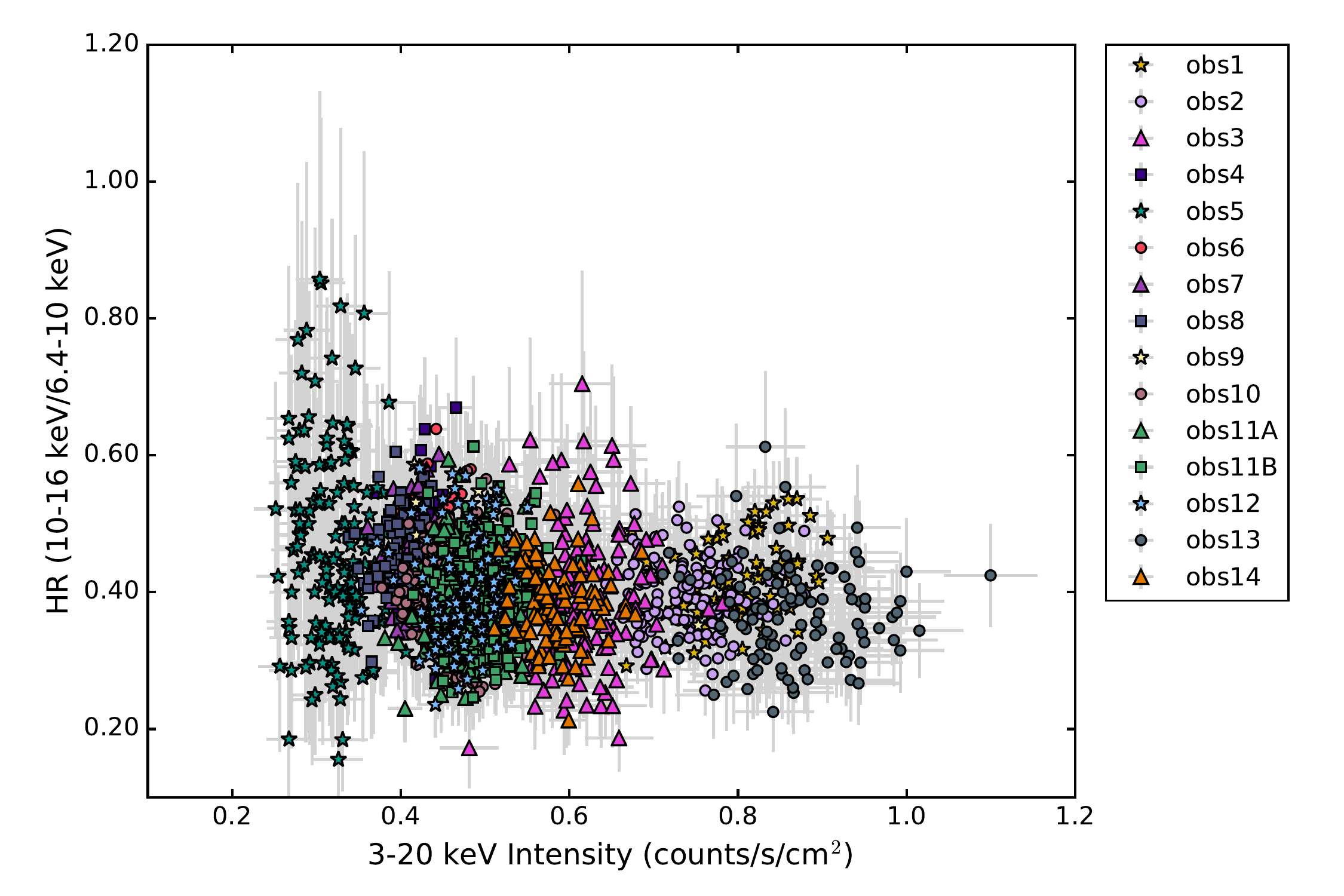}
\caption{Hardness-Intensity diagram of the straylight observations of  \gx. Observation numbers refer to the sequence IDs in Table \ref{tab:obs}. Data are binned to 300~s. The `banana' branch is traced out by the data. }
\label{fig:hid}
\end{center}
\end{figure}

\subsection{GX 3+1}

\gx is a persistently accreting `atoll' source. Atoll sources trace out regions on hardness-intensity diagrams that resemble `islands' (for which they are named: \citealt{hasinger_two_1989}) or `banana' shapes. \gx exclusively occupies the banana branch \citep{seifina_gx_2012} and was serendipitously observed via straylight in \nustar nineteen times between 2012 July and 2020 May. Table \ref{tab:obs} shows the sequence ID, observation date, FPM that the straylight occurred on, exposure time, and area on the FPM for observations with an area greater than 1~cm$^2$ of straylight from the source. Lightcurves were generated in three different energy bands ($3-20$~keV, $6.4-10$~keV, and $10-16$~keV) with a binsize of 300~s. Figure \ref{fig:hid} shows the hardness-intensity diagram for \gx. The hardness ratio (HR) is defined as the $10-16$~keV band divided by the $6.4-10$~keV band \citep{coughenour_nustar_2018}. The source traces out the `banana' branch.

\begin{figure}[!]
\begin{center}
\includegraphics[angle=270,width=0.45\textwidth,trim=0 40 0 0,clip]{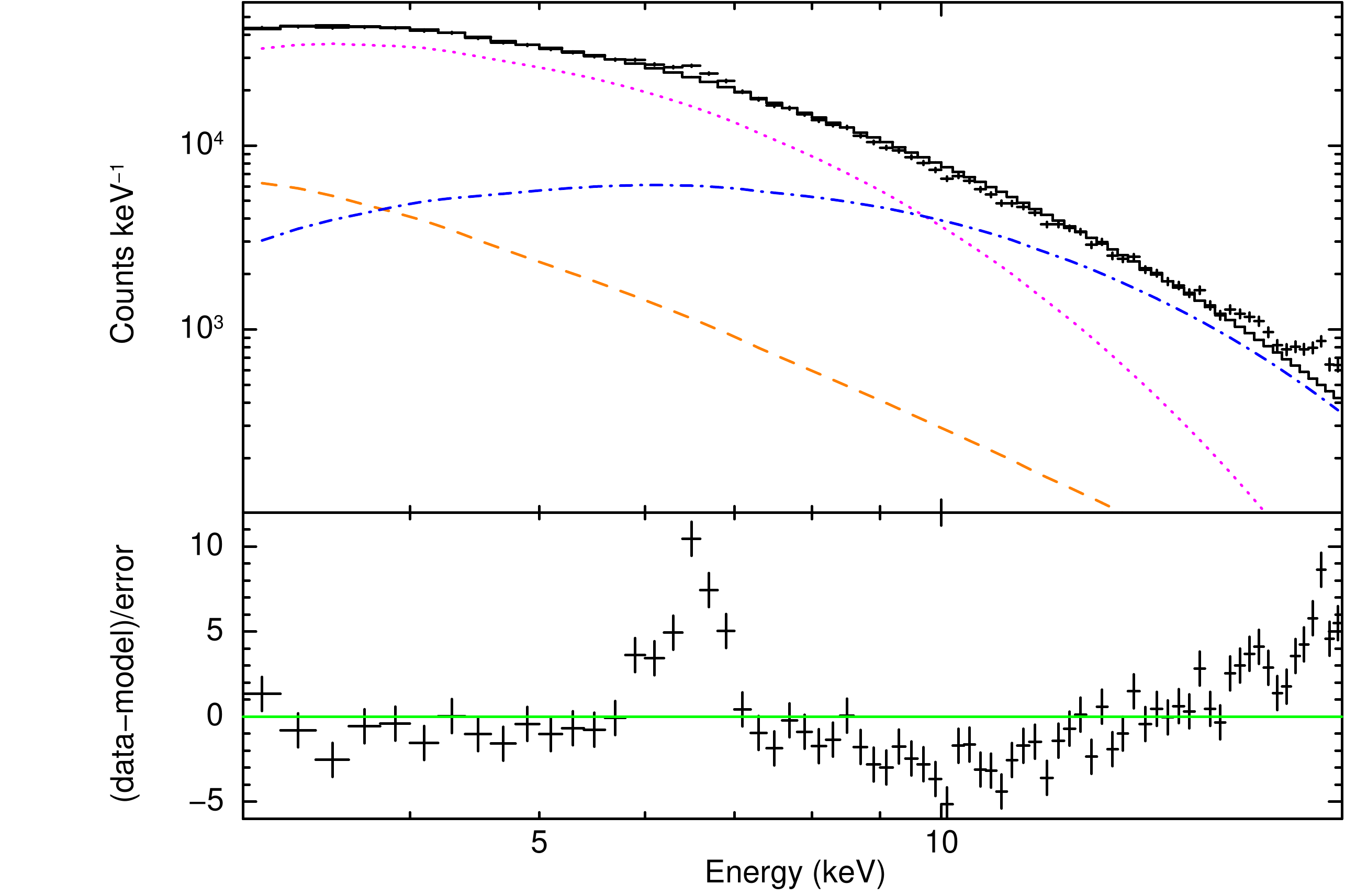}
\caption{The 3--20 keV straylight spectrum of \gx obs10 and residuals divided by the error. The orange dashed line indicates the power-law component, the blue dot-dashed line is the single-temperature blackbody, the dotted line is the multi-temperature blackbody. A prominent Fe line feature is present between $6-7$~keV. The background begins to dominate above 15~keV. }
\label{fig:model}
\end{center}
\end{figure}

To demonstrate the spectral utility of straylight observations for studying NS LMXBs, we extract a spectrum from the longest observation, obs10. The data are fit with the three component model of \citet{lin_evaluating_2007} that was used in \citet{ludlam_nustar_2019} for the pointed observation of \gx. This is comprised of a multi-temperature blackbody for thermal emission from the accretion disk, single-temperature blackbody for a boundary layer or emission from the NS surface, and power-law for weak Comptonized emission. For direct comparison to the intentional \nustar observation, we model the continuum emission by fixing the absorption column along the line of sight, blackbody temperatures, and photon index to the values reported in Table~2 of \citet{ludlam_nustar_2019} while allowing for the normalizations of each spectral component to vary. The spectrum and continuum components are shown in Figure \ref{fig:model}. The color scheme and line types correspond to those in \citet{ludlam_nustar_2019}. Indeed, a prominent Fe line emission feature can be seen in the straylight observations akin to the one observed from the pointed observations (see Fig~1 of \citealt{ludlam_nustar_2019}).  Further details of the variations in this source over time will be addressed in future work.

\begin{figure}[h!]
\begin{center}
\includegraphics[width=0.5\textwidth]{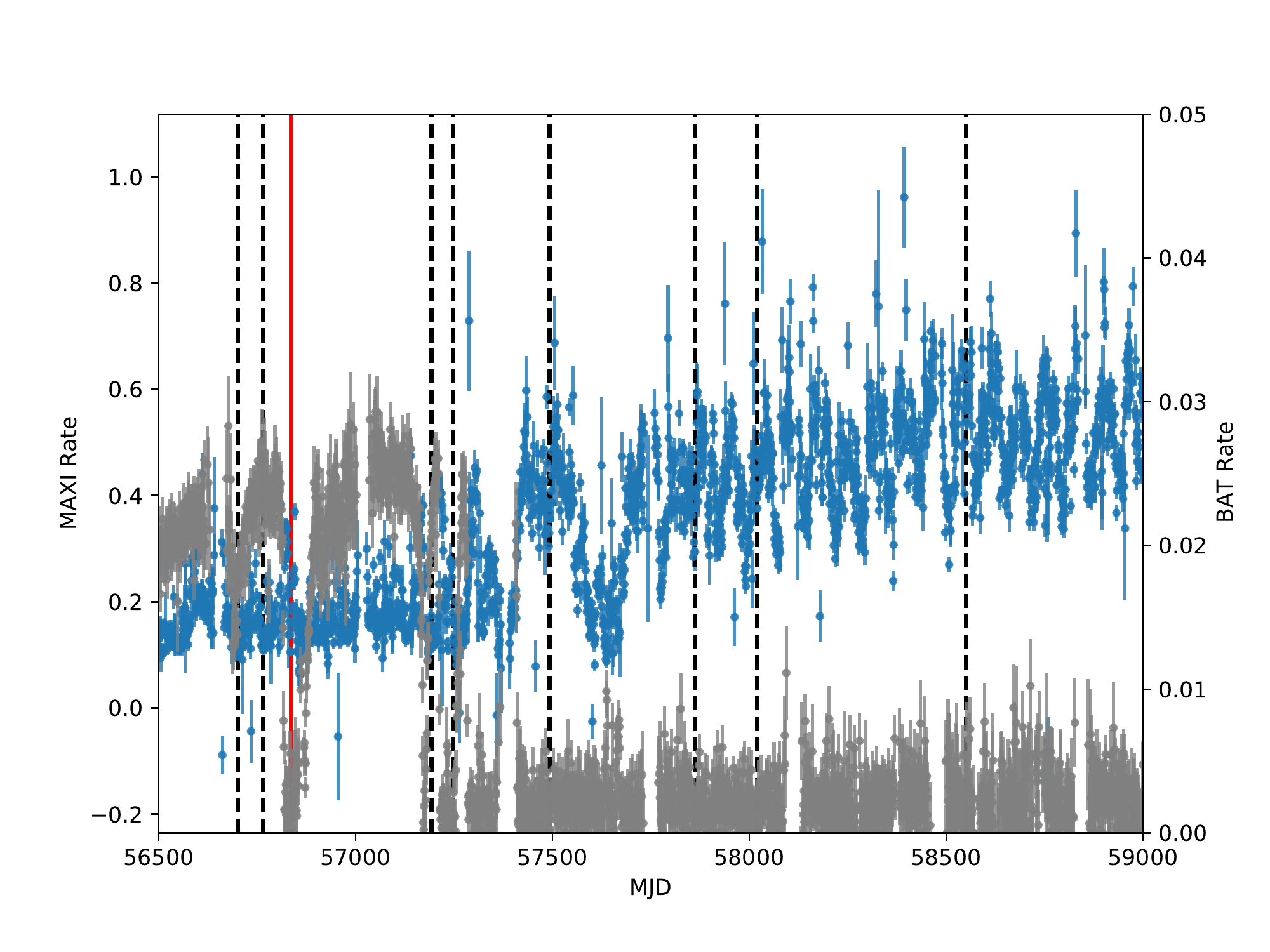}
\caption{\label{fig:gs1826longterm} The long-term 2--20 keV MAXI lightcurve (blue) and the \swift-BAT transient monitor 15-50 keV lightcurve for GS 1826-24 (grey). The timing of the focused \nustar observations are shown in solid red lines while the timing of the \straycats  observations are shown in dashed black lines.}
\end{center}
\end{figure}

\subsection{GS 1826-24}

GS 1826-24 is a LMXB which showed remarkable consistent Type I X-ray bursts since its discovery by \ginga\ \citep[e.g.][]{ubertini_bursts_1999}. The Type I X-ray bursts were so regular as to earn this source the ``Clocked Burster" moniker. A sudden dip in the \swift-BAT 15-50 keV lightcurve resulted in a \nustar ToO observation of this source in 2014 \citep{chenevez_soft_2016}. After briefly returning to a hard state, the source appears to have transitioned into a ``soft" state in 2016 with the MAXI lightcurve increasing to a plateau in 2018 and the \swift-BAT lightcurve in an apparently quiescent state (Fig \ref{fig:gs1826longterm}). While there have not been any subsequent targeted observations with either \nustar or \xmm, \nicer  has monitored the source and found evidence for mHz QPOs \citep{strohmayer_nicer_2018}.

\begin{figure*}[ht!]
\begin{center}
\includegraphics[width=0.49\textwidth]{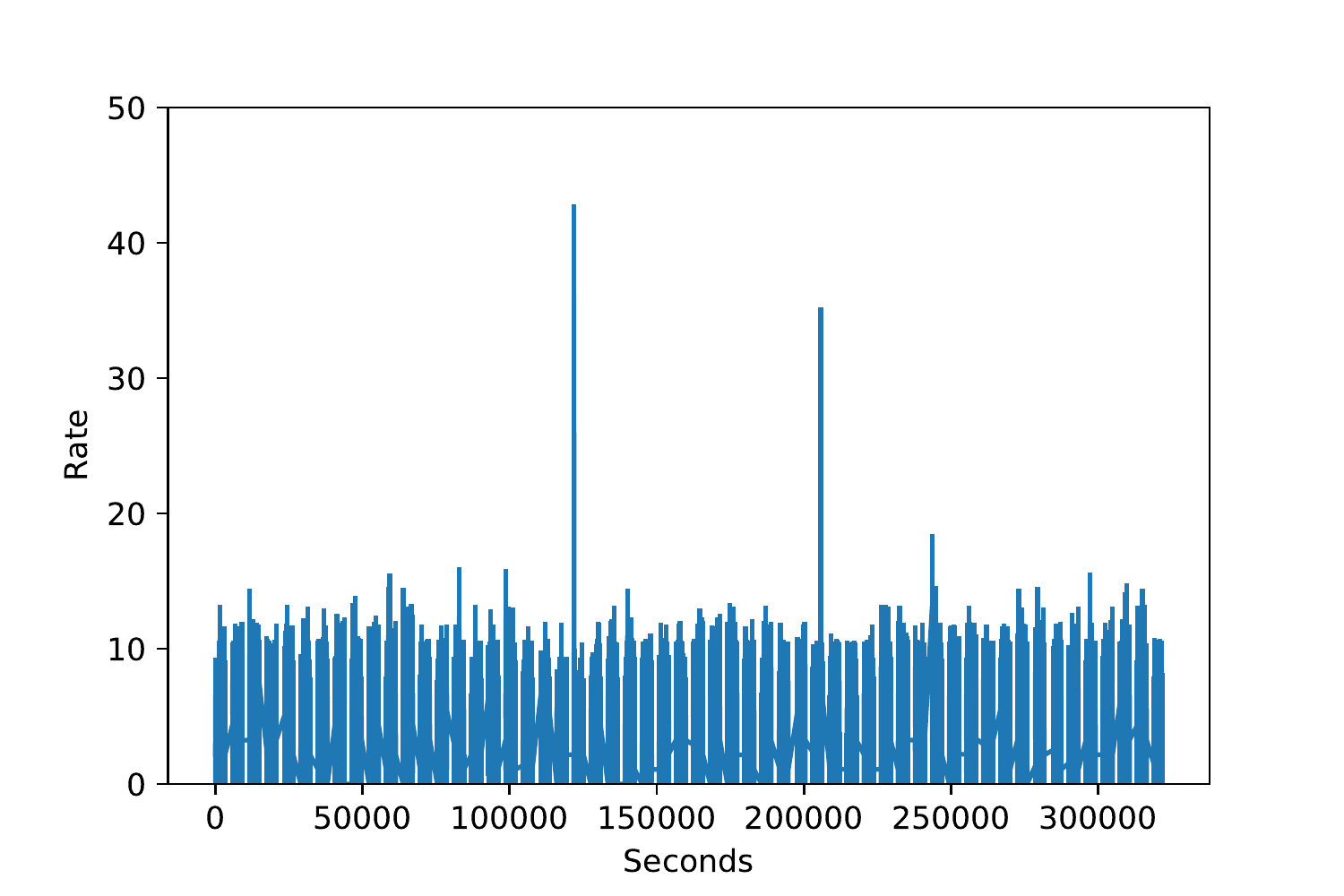} \includegraphics[width=0.49\textwidth]{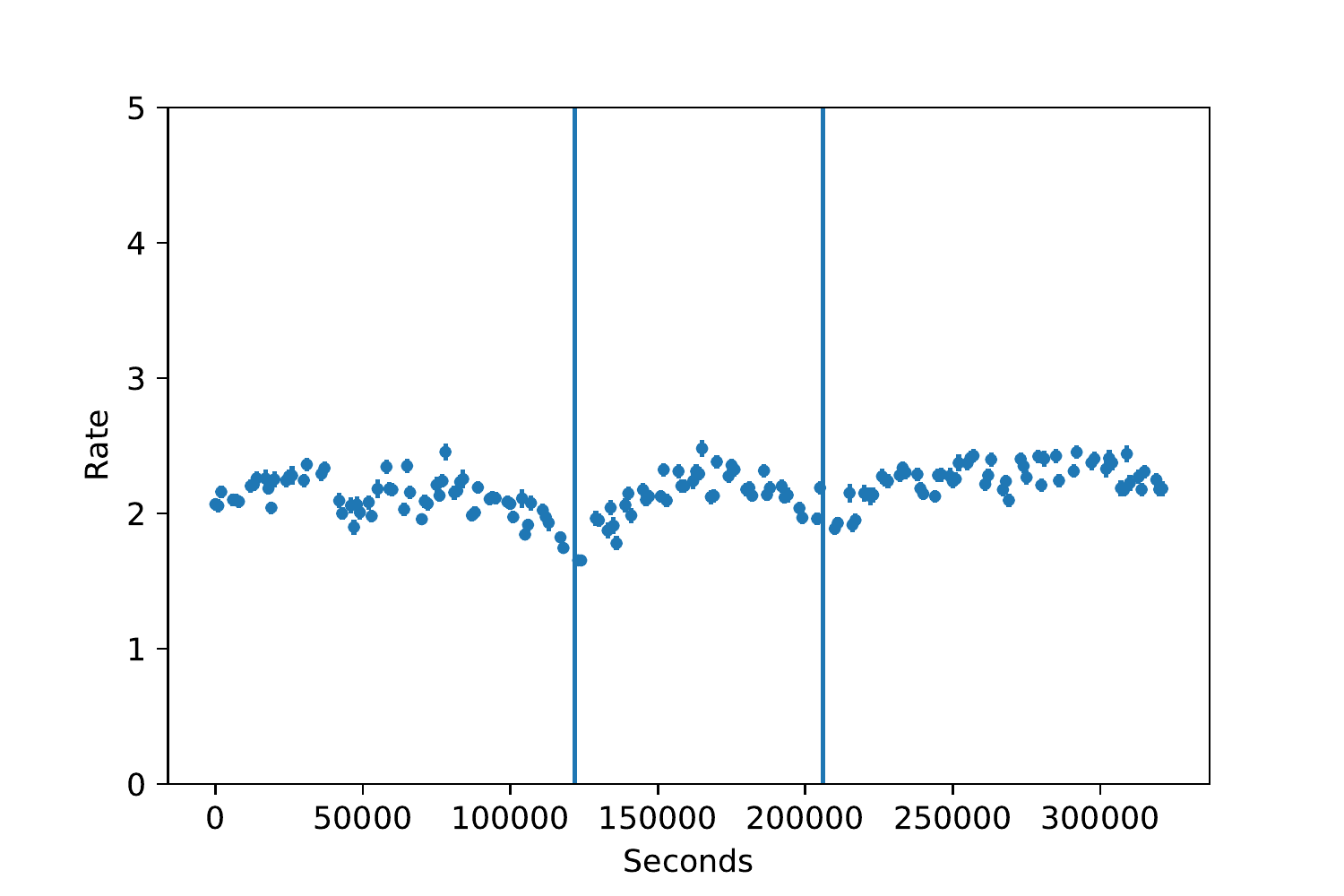} \\
\includegraphics[width=0.49\textwidth]{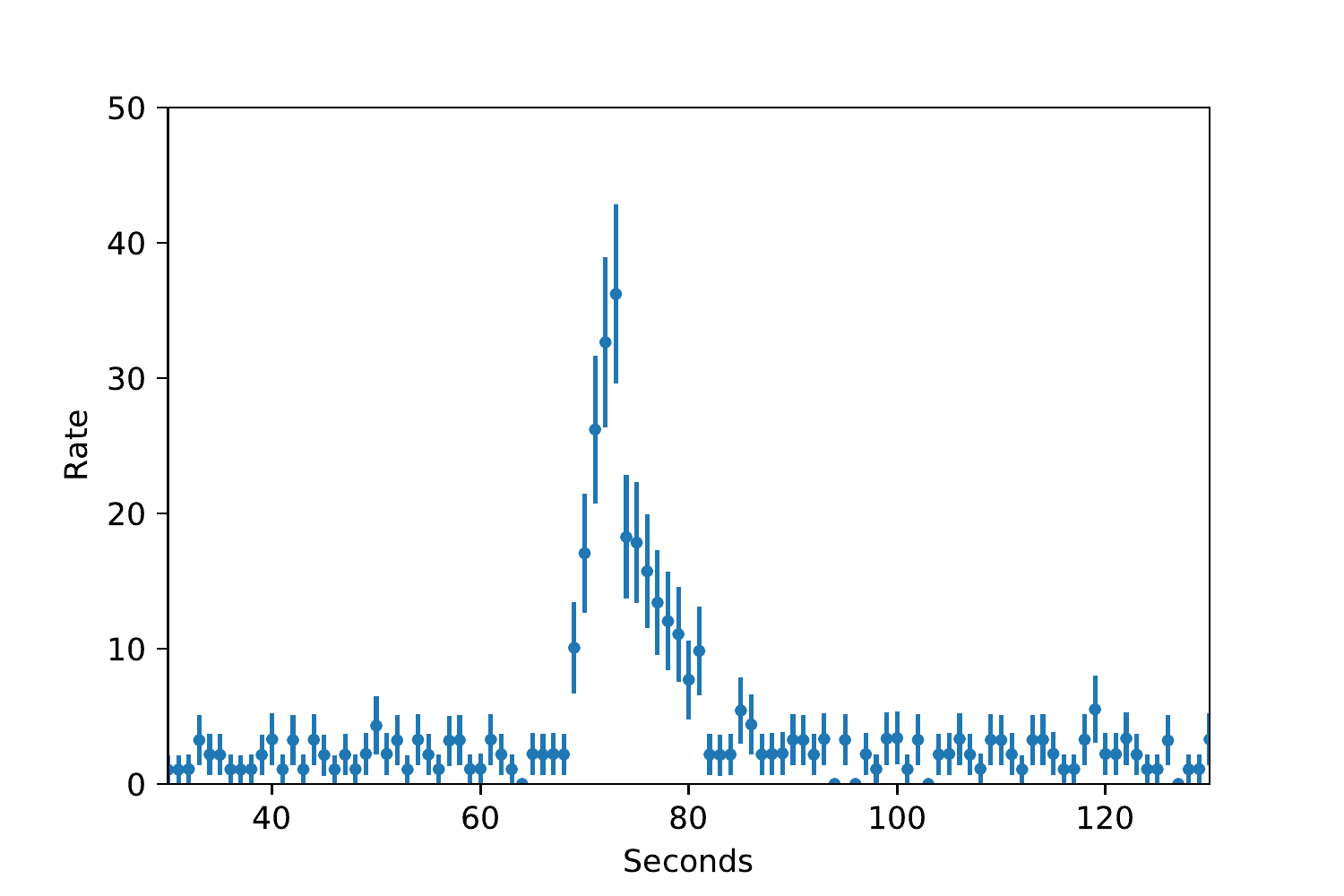} \includegraphics[width=0.49\textwidth]{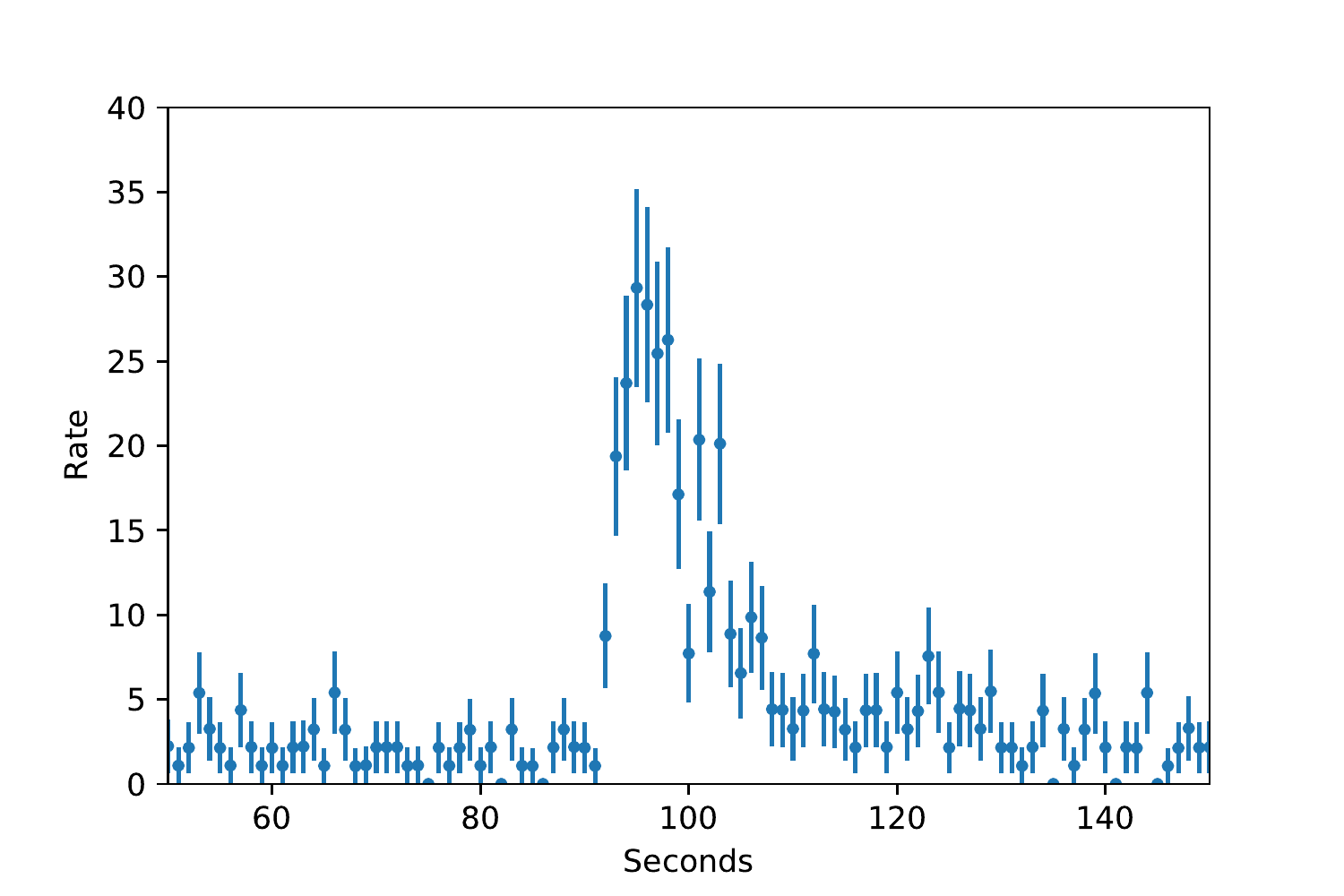}
\caption{\label{fig:gs1826lc} All panels show the 3--20 keV lightcurve of Obs7 and show: (\textit{Top Left}) The full observation using 1-s bins clearly shows the two Type I X-ray bursts; (\textit{Top Right}) The same data, but using 1-ks bins; (\textit{Bottom panels}) The zoomed in view of the first (\textit{left}) and second (\textit{right}) Type I X-ray burst.
}
\end{center}
\end{figure*}

The \straycats  observations (Table \ref{tab:gs1826obs}) span both the pre-dip observations and include several long observations during the BAT X-ray minimum. We highlight one of these (Obs7), which had a substantial amount of stray light covering over half of FPMB and a long exposure of over 150-ks, resulting in nearly 300-ks of elapsed clock time. During this observation \nustar clearly detected two Type I X-ray bursts lasting $\sim$10s of seconds (Fig \ref{fig:gs1826lc}). Simultaneously, the X-ray flux in the 3--20 keV lightcurve dipped leading up to the burst itself. We only find two Type I X-ray bursts, while we would have expected over a dozen had the source been regularly bursting with a recurrence time of $\sim$5.7 hours \citep{ubertini_bursts_1999}. This confirmed the results of the single set of pointed \nustar observations that the ``clocked" nature of the source has disappeared in the soft state \citep{chenevez_soft_2016}. A more complete survey of the bursting state over all 7 epochs and correlations with the spectral changes in the source will be the topic of a future paper.

\begin{table*}
\caption{GS 1826-24 \straycats  Observations}
\label{tab:gs1826obs}
\begin{center}
\begin{tabular}{lccccccc}
\hline
Obs \# & Sequence ID & Obs.\ Date &  (MJD) & FPM & Exp.\ (ks) & area (cm$^{2}$) \\
\hline
1 & 80002012002 & 2014-02-14T00:36:07.184 &  56702.0 & A &    24.05 & 2.2 \\
2 & 80002012004 & 2014-04-17T22:46:07.184 &  56764.9 & A &    26.42 & 2.3 \\
3 & 30101053002 & 2015-06-17T16:06:07.184 &  57190.7 & A &   131.32 & 2.75 \\
4 & 30101053004 & 2015-06-21T07:11:07.184 &  57194.3 & A &    51.52 & 2.5 \\
5 & 60160692002 & 2016-04-14T18:26:08.184 &  57492.8 & B &    21.88 & 1.7 \\
6 & 10202005002 & 2017-04-18T13:06:09.184 &  57861.5 & A &   156.51 & 2.52 \\ 
7* & 10202005004 & 2017-09-23T08:36:09.184 &  58019.4 & B &   156.54 & 8.8 \\
8 & 80460628002 & 2019-03-08T20:21:09.184 &  58550.8 & B &    41.39 & 1.6 \\
\hline
\end{tabular}
\end{center}
*:Used for the analysis in this work
\end{table*}

\section{Summary and future work}

In this paper we have presented a summary of a unique, untapped set of \nustar observations. The \straycats observations found thus far are predominantly associated with known bright sources and transient X-ray binaries as they go into outburst. 

\straycats is based on a systematic approach to mining the database of \nustar observations. While previously these observations were considered a nuisance, we have now produced a set of publicly available tools for analyzing these data and producing high-level science products. In addition, we provided access to scripts that help in the generation of region files. which often requires some fine tuning based on the projected ``shadow" of the optics bench.

The \straycats catalog that we present here we consider to be version 1.0. We intend to extend the current version of \straycats to include additional summary data products (such as count rates, hardness ratios, and source and background extraction regions) for all \straycats observations where the source is bright enough and enough of the focal plane is covered by stray light. This work is on-going and will be provided in a future release.

Finally, our brief survey of the science potential from \straycats  observations shows the power of these observations. Through these highlights of a few selected observations we have shown that these data can be used to track sources over long periods of time and provide a unique window into their behavior by providing improved sensitivity and finer spectral resolution compared to other all-sky monitors such as MAXI and \swift-BAT. 

\section*{Acknowledgements}

This work was supported by the National Aeronautics and Space Administration (NASA) under grant number 80NSSC19K1023 issued through the NNH18ZDA001N Astrophysics Data Analysis Program (ADAP).
R.M.L. acknowledges the support of NASA through Hubble Fellowship Program grant HST-HF2-51440.001. R.A.K. acknowledges support from the Russian Science Foundation (grant 19-12-00396).
JH acknowledges support from an appointment to the NASA Postdoctoral
Program at the Goddard Space Flight Center, administered by the USRA through a contract with NASA.

Additionally, this work made use of data from the \nustar mission, a project led by the California Institute of Technology, managed by the Jet Propulsion Laboratory, and funded by the National Aeronautics and Space Administration. We thank the \nustar Operations, Software and Calibration teams for support with the execution and analysis of these observations. This research has made use of the \nustar Data Analysis Software (NuSTARDAS) jointly developed by the ASI Science Data Center (ASDC, Italy) and the California Institute of Technology (USA). This research has made use of data and/or software provided by the High Energy Astrophysics Science Archive Research Center (HEASARC), which is a service of the Astrophysics Science Division at NASA/GSFC.

\facilities{NuSTAR,
            Swift,
            MAXI,
            HEASARC}

\software{astropy \citep{robitaille_astropy_2013, astropy_collaboration_astropy_2018},
          astroquery \citep{ginsburg_astroquery_2019},
          HEASoft/FTOOLS,
          IDL\footnote{\url{https://www.harrisgeospatial.com/Software-Technology/IDL}},
          matplotlib\citep{hunter_matplotlib_2007},
          numpy\footnote{\url{https://numpy.org}},
          pandas\citep{pandas_development_team_pandas-devpandas_2020},
          plotly\footnote{\url{https://plotly.com/python/}},
          scikit-image\footnote{\url{https://scikit-image.org}},
          Veusz\footnote{\url{https://veusz.github.io}}
          }

\appendix 

\begin{longrotatetable}
\begin{deluxetable*}{llccccccccccc}

\tablecaption{\straycats Excerpt \label{tab:straycats}}
\tablewidth{700pt}
\tabletypesize{\scriptsize}
\tablehead{
\colhead{STRAYID} & \colhead{Classif.} & 
\colhead{SEQID} & \colhead{Mod.} & 
\colhead{Primary} & \colhead{TSTART} & 
\colhead{Exp} & \colhead{SL Source} & 
\colhead{SL Type} & \colhead{RA$_{SL}$} &
\colhead{DEC$_{SL}$}  & \colhead{RA$_{Pri}$} &
\colhead{DEC$_{Pri}$}
} 
\startdata
StrayCatsI\_0 & Faint & 30001014002 & B & IC10\_X1 &  56967.8 &    88.47 & NA & ?? & -999 & -999 &    5.074 &   59.274  \\ 
StrayCatsI\_1 & Unkn & 90101010002 & A & IGR\_J00291p5934 &  57231.5 &    38.10 & NA & ?? & -999 & -999 &    7.275 &   59.598  \\ 
StrayCatsI\_2 & SL & 90201030002 & A & SWIFT\_J003233d6m7306 &  57586.7 &    54.92 & SMC X-1 & HMXB-NS &   19.271 &  -73.443 &    8.197 &  -73.096  \\ 
StrayCatsI\_3 & SL & 90201041002 & A & SMC\_X3 &  57704.7 &    38.60 & SMC X-1 & HMXB-NS &   19.271 &  -73.443 &   13.030 &  -72.457  \\ 
StrayCatsI\_4 & SL & 30361002002 & A & SXP\_15d3 &  58087.0 &    70.65 & SMC X-1 & HMXB-NS &   19.271 &  -73.443 &   13.112 &  -73.373  \\ 
StrayCatsI\_5 & SL & 30361002004 & A & SXP\_15d3 &  58418.9 &    58.35 & SMC X-1 & HMXB-NS &   19.271 &  -73.443 &   13.141 &  -73.350  \\ 
StrayCatsI\_6 & Faint & 30361002004 & B & SXP\_15d3 &  58418.9 &    58.13 & NA & ?? & -999 & -999 &   13.141 &  -73.350  \\ 
StrayCatsI\_7 & Faint & 50311003002 & A & SMC\_Deep\_MOS03 &  57876.1 &   172.39 & NA & ?? & -999 & -999 &   13.264 &  -72.488  \\ 
StrayCatsI\_8 & SL & 60301029006 & A & IRAS\_00521m7054 &  58074.7 &    74.41 & SMC X-1 & HMXB-NS &   19.271 &  -73.443 &   13.498 &  -70.667  \\ 
StrayCatsI\_9 & SL & 90101017002 & A & SMC\_X2 &  57316.9 &    26.72 & SMC X-1 & HMXB-NS &   19.271 &  -73.443 &   13.740 &  -73.707  \\ 
StrayCatsI\_10 & SL & 90102014004 & A & SMC\_X2 &  57307.9 &    23.05 & SMC X-1 & HMXB-NS &   19.271 &  -73.443 &   13.744 &  -73.695  \\ 
StrayCatsI\_11 & SL & 50311001002 & B & SMC\_Deep\_MOS01 &  57867.1 &   153.16 & SMC X-1 & HMXB-NS &   19.271 &  -73.443 &   13.930 &  -72.439  \\ 
StrayCatsI\_12 & SL & 30202004006 & B &  SMC\_X1 &  57662.0 &    20.38 & SMC X-3 & HMXB-NS &   13.023 &  -72.435 &   19.351 &  -73.462  \\ 
StrayCatsI\_13 & SL & 30202004002 & A & SMC\_X1 &  57639.9 &    22.46 & RX J0053.8-7226 & HMXB-NS &   13.480 &  -72.446 &   19.378 &  -73.442  \\ 
StrayCatsI\_14 & SL & 30202004002 & B & SMC\_X1 &  57639.9 &    22.53 & RX J0053.8-7226 & HMXB-NS &   13.480 &  -72.446 &   19.378 &  -73.442  \\ 
StrayCatsI\_15 & SL & 30202004004 & B & SMC\_X1 &  57650.3 &    21.18 & RX J0053.8-7226 & HMXB-NS &   13.480 &  -72.446 &   19.380 &  -73.454  \\ 
StrayCatsI\_16 & SL & 90001008002 & A & GK\_Per &  57116.1 &    42.35 & NGC 1275 & AGN &  49.951 & 41.512 &   52.784 &   43.870  \\ 
StrayCatsI\_17 & SL & 30101021002 & A & GK\_Per &  57273.7 &    72.30 & NGC 1275 & AGN &   49.951 &   41.512 &   52.812 &   43.933  \\ 
StrayCatsI\_18 & SL & 30101021002 & B & GK\_Per &  57273.7 &    72.16 & NGC 1275 & AGN &   49.951 &   41.512 &   52.812 &   43.933  \\ 
StrayCatsI\_19 & SL & 10601407002 & A & N132D &  58921.4 &    82.69 & LMC X-2 & LMXB-NS &   80.117 &  -71.965 &   81.197 &  -69.695  \\ 
StrayCatsI\_20 & SL & 40101010002 & B & N132D &  57366.2 &    68.85 & LMC X-4 & NS &   83.206 &  -66.370 &   81.311 &  -69.666  \\ 
StrayCatsI\_21 & SL & 40101010002 & B & N132D &  57366.2 &    68.85 & LMC X-2 & LMXB-NS &   80.117 &  -71.965 &   81.311 &  -69.666  \\ 
StrayCatsI\_22 & SL & 30460020002 & A & 1RXS\_J052523d2p241331 &  58563.6 &    58.79 & Crab & PWNe &   83.633 &   22.015 &   81.316 &   24.192  \\ 
StrayCatsI\_23 & SL & 30460020002 & B & 1RXS\_J052523d2p241331 &  58563.6 &    58.29 & Crab & PWNe &   83.633 &   22.015 &   81.316 &   24.192  \\ 
StrayCatsI\_24 & SL & 30301014002 & A & SGR\_0526m66 &  58156.9 &    47.03 & LMC X-3 & LMXB-BH &   84.736 &  -64.084 &   81.500 &  -66.100  \\ 
StrayCatsI\_25 & SL & 30301014002 & B & SGR\_0526m66 &  58156.9 &    46.90 & LMC X-3 & LMXB-BH &   84.736 &  -64.084 &   81.500 &  -66.100  \\ 
\enddata
\end{deluxetable*}
\end{longrotatetable}

\bibliographystyle{aasjournal}

\bibliography{refs.bib}

\end{document}